\documentclass[journal]{IEEEtran}
\usepackage{amssymb}
\usepackage{graphicx}
\usepackage{caption}
\usepackage{subfigure}
\usepackage{amsmath}
\usepackage{float}
\usepackage{ragged2e}
\usepackage{color}
\usepackage{epstopdf}
\usepackage{lineno,hyperref}
\usepackage[figuresright]{rotating}
\usepackage{xcolor}
\usepackage{bm}
\usepackage{xpatch}
\usepackage{setspace}
\usepackage{balance}
\modulolinenumbers[5]
\newtheorem{assumption}{\textbf{Assumption}}

\newtheorem{theorem}{\textbf{Theorem}}
\newtheorem{proposition}{\textbf{Proposition}}
\newtheorem{definition}{\textbf{Definition}}
\newtheorem{lemma}{\textbf{Lemma}}
\newtheorem{remark}{\textbf{Remark}}

\newcommand{\tabincell}[2]{\begin{tabular}{@{}#1@{}}#2\end{tabular}}
\usepackage[justification=centering]{caption}
\usepackage{booktabs}
\usepackage{multirow}
\usepackage{cite}
\usepackage{makecell}
\hypersetup{
colorlinks=true,
linkcolor=black,
citecolor=black,
anchorcolor=black
}

\usepackage{fancyhdr}
\providecommand{\keywords}[1]{\textbf{\textit{Index terms---}} #1}
\begin{document}

\title{Distributed Control of Multi-agent Systems with Unknown Time-varying Gains: A Novel Indirect Framework for Prescribed Performance}

\author{Zeqiang Li,
        Yujuan~Wang,
        Xiucai~Huang$^*$
\thanks{This work was supported in part by the National Natural Science Foundation of China under Grant 61991400, 61991403, 61860206008, 61933012, and in part by the Natural Science Foundation of Chongqing cstc2019jcyj-msxmx0319.}
\thanks{The authors are with the State Key Laboratory of Power Transmission Equipment System Security and New Technology, Chongqing Key Laboratory of Intelligent Unmanned Systems, School of Automation, Chongqing University, Chongqing 400044, China, and also with the Star Institute for Intelligent Systems, Chongqing 400044, China (e-mail:iamlzq@cqu.edu.cn; yjwang66@cqu.edu.cn; hxiucai@cqu.edu.cn).}
}

\markboth{}{ \MakeLowercase{\textit{et al.}}: }
\maketitle

\begin{abstract}
In this paper, a new yet indirect performance guaranteed framework is established to address the distributed tracking control problem for networked uncertain nonlinear strict-feedback systems with unknown time-varying gains under a directed interaction topology. The proposed framework involves two steps: In the first one, a fully distributed robust filter is constructed to estimate the desired trajectory for each agent with guaranteed observation performance that allows
the directions among the agents to be non-identical.
In the second one, by establishing a novel lemma regarding Nussbaum function, a new adaptive control protocol is developed for each agent based on backstepping technique, which not only steers the output to asymptotically track the corresponding estimated signal with arbitrarily prescribed transient performance, but also largely extends the scope of application since the unknown control gains are allowed to be time-varying and even state-dependent. In such an indirect way, the underlying problem is tackled with the output tracking error converging into an arbitrarily pre-assigned residual set exhibiting an arbitrarily pre-defined convergence rate. Besides, all the internal signals are ensured to be semi-globally ultimately uniformly bounded (SGUUB). Finally, simulation results are provided to illustrate the effectiveness of the co-designed scheme.
\end{abstract}

\begin{keywords}
Multi-agent systems, unknown time-varying gains, prescribed performance, distributed tracking.
\end{keywords}

\IEEEpeerreviewmaketitle

\section{Introduction}
\allowdisplaybreaks
Distributed tracking control of uncertain nonlinear multi-agent systems (MASs) plays a pivotal role in the control community since it enables each agent to operate in a desired way just by using local information. During the past decades, considerable research efforts have been devoted in this direction and significant advance has been accordingly achieved through adaptive backstepping \cite{shen2015distributed, su2015cooperative, li2020distributed} and adaptive neural network (NN)/fuzzy logic control \cite{zhang2012adaptive, liang2020event}, exhibiting its theoretical importance and broad practical application prospect.

The current related results are commonly built upon the prior knowledge of control direction of each agent, which might be not the case in practice. For instance, it would be non-trivial to explore the control directions of robotic visual servo systems \cite{jiang2002iterative} and ship autopilot systems \cite{du2007adaptive}.
To tackle such problem, there are mainly three systematic technologies: Nussbaum function \cite{nussbaum1983some, xudong1998adaptive}, extremum seeking \cite{oliveira2011output, scheinker2012minimum} and logic switching mechanism \cite{wu2016global, huang2018tuning}, among which the Nussbaum function serves as the predominant one. In \cite{chen2013adaptive, liu2014adaptive, ding2015adaptive}, a set of identical Nussbaum functions are utilized to handle unknown direction control problem for uncertain MASs, where the control directions are required to be identical to circumvent the counteract of Nussbaum gains in the stability analysis. Subsequently, significant advances for the case with non-identical control directions have also been made in
\cite{chen2016adaptive, psillakis2016consensus, wang2018cooperative, wang2020adaptive}. By developing two kinds of Nussbaum function, the limitation of identical control directions among agents is relaxed in \cite{chen2016adaptive}, yet only part of such non-identical directions are allowed to be unknown. In \cite{psillakis2016consensus, wang2018cooperative}, the PI consensus error is incorporated into the control design to remove this assumption while solving leaderless consensus control problem, which thereby are invalid for distributed tracking problem.
Parallelly, the authors in \cite{wang2020adaptive} resolve such issue by resorting to a series of different Nussbaum functions, which inevitably makes the stability analysis rather complicated. Notably, the control gains (also referred to as control coefficients in \cite{chen2019nussbaum}) in \cite{chen2013adaptive, liu2014adaptive, ding2015adaptive, psillakis2016consensus, wang2018cooperative, wang2020adaptive} are limited to be constant instead of time-varying, thus largely restricting the scope of application. Some attempts are made in \cite{huang2018fully, wang2020adaptiveb} to break through this restriction, yet the time-varying part of the coefficients is required to be known. Therefore, a natural motivation of this paper is how to achieve distributed tracking control of uncertain nonlinear MASs in the presence of completely unknown time-varying coefficients with non-identical signs.

Apart from the unknown control direction problem, performance metrics that characterize the transient and steady-state properties of the tracking error (i.e., the convergence rate and the size of steady-state error) are also vital considerations in the control design for uncertain nonlinear MASs (see \cite{verginis2019robust, mehdifar2020prescribed} for examples).
One typical method to achieve such goal is to combine the so-called prescribed performance control with NN/fuzzy approximation technique, see \cite{liang2019prescribed, meng2019distributed} for instance, which inevitably suffers from heavy computational burden due to the involvement of highly complex approximation structure.
Some approximation-free prescribed performance control methods with low complexity are developed in \cite{bechlioulis2016decentralized, katsoukis2021low} for uncertain nonlinear MASs in Brunovsky canonical/triangular forms, respectively.
Working independently, the authors in \cite{huang2022distributed} propose a distributed performance-guaranteed control strategy for multi-inputs multi-outputs (MIMO) nonlinear MASs under some relaxed controllability conditions. However, the control directions in \cite{verginis2019robust, mehdifar2020prescribed, liang2019prescribed, meng2019distributed, bechlioulis2016decentralized, katsoukis2021low, huang2022distributed} are uniformly assumed to be known \emph{a priori}.
Although some approaches are proposed to tackle this problem using orientation functions, they are applicable only for uncertain nonlinear single systems (see \cite{zhang2019low, zhang2021global}), it thus remains unclear whether such methods can be extended to MASs.
Hence, another motivation of this paper is how to achieve prescribed performance control of uncertain nonlinear MASs with unknown control directions.

Inspired by such observations, in this work we investigate the distributed performance guaranteed problem for a class of networked uncertain nonlinear strict-feedback systems with unknown and non-identical control directions under a directed protocol, where the unknown coefficients in the input channel are allowed to be time-varying and even state-dependent, which is thus more general than the systems considered in \cite{chen2013adaptive, wang2020adaptive, liu2014adaptive, ding2015adaptive, chen2016adaptive, psillakis2016consensus, wang2018cooperative}. It is a non-trivial task to achieve prescribed tracking performance for such kind of systems since the involved uncertainties (including mismatched parametric uncertainties and unknown input gains) tend to degrade the system performance and the coupling of the non-identical control directions among the agents would challenge both control design and stability analysis. In order to tackle those issues, a novel performance guaranteed framework is constructed, which is comprised of two steps: In the first step, a fully distributed performance guaranteed filter is designed to reconstruct the desired trajectory (i.e., the output of the leader) for each agent. The proposed design is not only able to steer the filter error into an arbitrarily pre-assignable residual set with an arbitrarily pre-set converge rate, but also able to decouple the problem of unknown non-identical control directions among agents, making it more powerful than the ones in \cite{wang2020adaptiveb, huang2017smooth} since their transient observation performance cannot be explicitly prescribed. Besides, our structure remains at a lower complexity level as no extra adaptive parameters are required to be updated online; In the second step, by incorporating the desired performance characteristics into the backstepping design procedure, an adaptive control scheme is proposed for each agent, which, on one hand, enables the output to asymptotically track the corresponding estimated signal with arbitrarily prescribed transient performance, on the other hand, is able to address the unknown control direction problem for a wider set of systems with time-varying and even state-dependent coefficients by resorting to a newly casted lemma regarding Nussbaum functions. Based on such an indirect framework, it is eventually guaranteed that the output of each agent follows the desired trajectory with arbitrarily pre-specified transient and steady-state performance, i.e., with arbitrarily pre-assignable converge speed and arbitrarily prescribed size of the residual set. Finally, all the internal signals are ensured to be SGUUB and sufficient simulation studies are carried out to illustrate the effectiveness and benefits of such co-designed scheme.

The rest of this article is set up as follows. Section II provides some preliminaries and problem statement. Next, Section III displays the main results including control design and stability analysis. In Section IV, we conduct the numerical simulation and then Section V concludes the paper.

\emph{Notations}: Let $\Re^n$ be the set of real vectors of order $n$. Bold notations stand for matrices (or vectors). $\|\cdot\|$ represents the standard Euclidean
norm. For a nonsingular matrix $\boldsymbol M \in \Re^{n \times n}$, $\lambda_{\mathrm{min}}\{\boldsymbol M\}$ and $\sigma_{\mathrm{min}}\{\boldsymbol M\}$ denote its minimum eigenvalue and minimum singular value, respectively. $\boldsymbol 1_n \in \Re^n$ signifies a vector of ones.

\emph{Graph Theory}: Suppose that the networked topology among the followers is expressed by a directed graph ${\cal G} = \left( {{\cal V},{\cal E}} \right)$, in which ${\cal V} = \left\{ {1,...,N} \right\}$ is the set of vertices referring to the $N$ agents, and ${\cal E} \subseteq {\cal V} \times {\cal V}$ denotes the edge set of the graph. If $(i,j) \in \cal E$, then agent $j$ can receive information from agent $i$, and we say agent $i$ is a neighbor of agent $j$. In such condition, the neighboring set of node $i$ is denoted as ${{\cal N}_i}$. A sequence of edges connecting a sequence of nodes in the same direction is called a path in a graph. If there exists at least one node which has directed paths to all the other nodes in the directed graph, then it is said to have a directed spanning tree. Additionally, $\boldsymbol{A} = \left[ {{a_{ij}}} \right] \in {\Re^{N \times N}}$ denotes the connectivity matrix of graph $\cal{G}$, where $a_{ij}>0$ if $(j,i) \in {\cal E}$; $a_{ij}=0$, otherwise. In this work, self loops are not allowed (i.e., $a_{ii}=0$). We introduce the in-degree matrix $\boldsymbol{D}= {\rm diag}({d}_1,...,{d}_{\emph N}) \in {\Re^{\emph N \times \emph N}}$ with ${d}_i=\sum\nolimits_{j \in {{\cal N}_i}} {{a_{ij}}}$. Then, the Laplacian matrix is defined as $\boldsymbol{ L}=\left[ {{l_{ij}}} \right] = \boldsymbol{D}-\boldsymbol{A}$. In addition, we use $b_i=1$ to indicate that the $i$th agent has access to the state information of the leader, otherwise $b_i=0$. The augmented graph is denoted as ${\bar{\cal G}} = \left( {{\bar{\cal V}},{\bar{\cal E}}} \right)$ with ${\bar{\cal V}} = \left\{ {0,1,...,N} \right\}$ and ${\bar{\cal E}} \subseteq {\bar{\cal V}} \times {\bar{\cal V}}$.

\section{Preliminaries and Problem Statement}

\subsection{An Indirect Performance-guaranteed Framework}
Consider a generalized tracking error $e(t) \in \Re$. By prescribed performance control \cite{bechlioulis2008robust}, it is referred to the scenario that the trajectory of $e(t)$ is confined within some quantified boundaries that can characterize arbitrarily fast decaying rate and arbitrarily small steady-state value. Mathematically, such evolution can be formulated as
\begin{equation}
|e(t)|<\beta(t), \quad\forall t\geq 0 \nonumber
\end{equation}
where $\beta(t):[0, \infty) \rightarrow \Re^+$ is referred to as the performance function, which is equipped with the following properties:
\begin{itemize}
\item[${\cal P}_1)$]
$\beta(t)$ is strictly positive and monotonically decreasing;
\end{itemize}
\begin{itemize}
\item[${\cal P}_2)$]
$\mathop {\lim }\limits_{t \to \infty } \beta(t)=\beta_\infty > 0$;
\end{itemize}
\begin{itemize}
\item[${\cal P}_3)$]
$\beta^{(k)}(t) ~(k=0,1,...,n)$ is bounded and piecewise continuous.
\end{itemize}
The performance functions in the family of prescribed performance control are typically chosen as hyperbolic tangent function \cite{meng2019distributed, zhang2021global} and exponential function \cite{verginis2019robust, mehdifar2020prescribed, liang2019prescribed, bechlioulis2016decentralized, katsoukis2021low, huang2022distributed, zhang2019low, bechlioulis2008robust}. In this work, the later is considered which takes the uniform form as $\beta(t)=(\beta_0-\beta_\infty){\rm exp}(-\iota t)+\beta_\infty$, where $\iota$ is a positive constant, $\beta_0$ and $\beta_\infty$ are finite design parameters such that $\beta_0>\beta_\infty>0$.

\begin{figure}[tbp]
  \centering
  \includegraphics[height=0.4\textwidth]{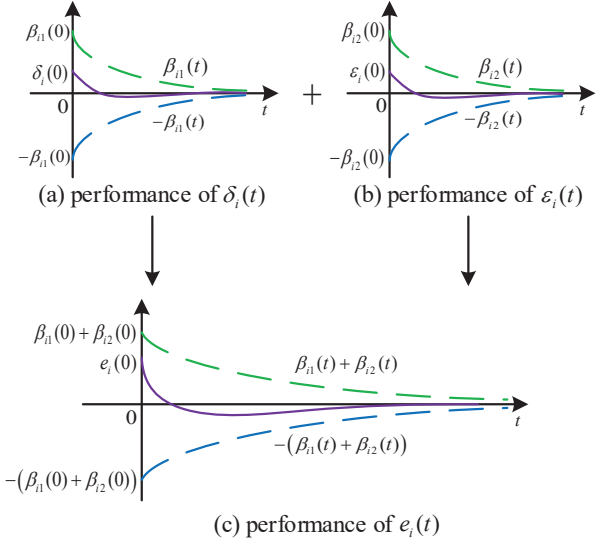}
  \caption{The relationship of performance bounds for $\delta_i$, $\varepsilon_i$ and $e_i$.}
  \label{Fig.1}
\end{figure}

Different from the centralized control strategy, in the distributed control for MASs, only part of the followers can access the desired trajectory $y_0(t)$, i.e., the output of the leader, making it significantly challenging to achieve prescribed performance for the output tracking error $e_i(t)$ directly, especially when the models of the agents suffer from strong nonlinearities and uncertainties. Motivated by the ongoing studies in \cite{wang2020adaptiveb, huang2017smooth, feng2016distributed}, an alternative is to design a distributed filter for each agent to estimate such desired signal, then $e_i(t)$ can be expressed as
\begin{align}\label{eq.2}
e_i(t) = \underbrace {{y_i(t)} - {{\hat y}_i(t)}}_{{\varepsilon_i(t)}} + \underbrace {{{\hat y}_i(t)} - {y_0(t)}}_{{\delta_i(t)}},
\end{align}
where $y_i(t)$ is the output of the $i$th agent, $\hat y_i(t)$ is the filtered variable, $\delta_i(t)$ and $\varepsilon_i(t)$ denote the filter error and the auxiliary tracking error, respectively. Intuitively, if the trajectories of $\delta_i(t)$ and $\varepsilon_i(t)$ are always preserved within some preset bounds that also are characterized by some performance functions $\beta_{i1}(t)$ and $\beta_{i2}(t)$, respectively, that is
\begin{align}
\label{eq.54}
|\delta_i(t)| &< \beta_{i1}(t)=(\beta_{i1,0}-\beta_{1,\infty})e^{-\iota t}+\beta_{1,\infty}, ~\forall t \geq 0, \\
\label{eq.52}
|\varepsilon_i(t)| &< \beta_{i2}(t)=(\beta_{i2,0}-\beta_{2,\infty})e^{-\iota t}+\beta_{2,\infty}, ~\forall t \geq 0,
\end{align}
it is immediate to obtain that
\begin{align}\label{eq.22}
|e_i(t)| \leq |\delta_i(t)|+ |\varepsilon_i(t)| < \beta_{i1}(t) + \beta_{i2}(t)=\beta_i(t),
\end{align}
for all $t \geq0$, where $\beta_i(t)$ is the performance function of $e_i(t)$. Thus, some explicit performance bounds for $e_i(t)$ can be calculated indirectly by $\beta_{i1}(t)$ and $\beta_{i2}(t)$, in the sense that the convergence speed is dictated by the parameter $\iota$ with larger $\iota$ leading to faster convergence speed, and the steady-state error is dominated by the parameter $\beta_{\ast,\infty}~(\ast=1,2)$ with smaller $\beta_{\ast,\infty}$ resulting in smaller error at steady state. The relationship of performance bounds for $\delta_i$, $\varepsilon_i$ and $e_i$ is shown in Fig. \ref{Fig.1}.

\subsection{Problem Formulation}
\begin{figure*}[htbp]
  \centering
  \includegraphics[height=0.24\textwidth]{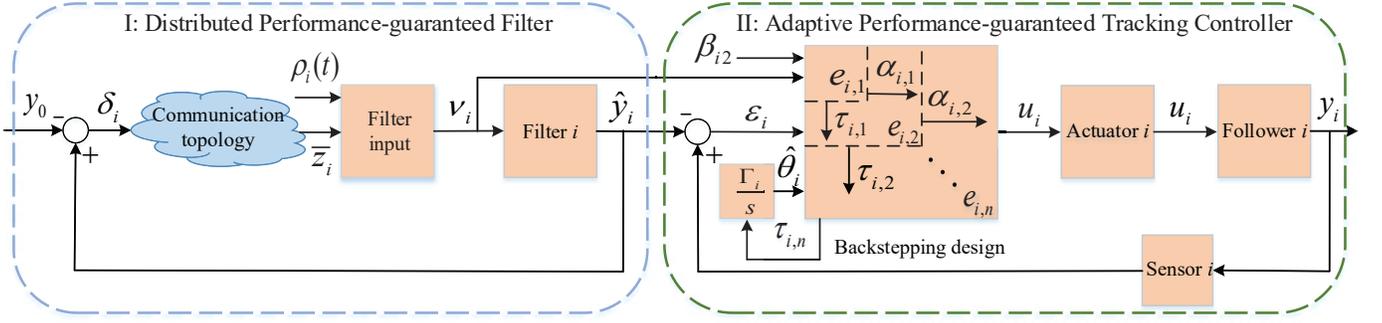}
  \caption{Block diagram of the proposed control algorithm for agent $i$.}
  \label{Fig.6}
\end{figure*}
Consider a group of networked systems consisting of $N$ agents that evolve according to
\begin{align}\label{eq.15}
{\dot x_{i,k}} &=x_{i,k+1}+\boldsymbol\theta^T_i\boldsymbol\varphi_{i,k}(x_{i,1},...,x_{i,k}), \quad k=1,...,n-1 \nonumber\\
{\dot x_{i,n}} &=g_i(\boldsymbol x_i,t)u_{i}+\boldsymbol\theta^T_i\boldsymbol\varphi_{i,n}(\boldsymbol x_i), \nonumber\\
y_i &=x_{i,1},   \quad i=1,...,N
\end{align}
where $\boldsymbol x_i=[x_{i,1},...,x_{i,n}]^T \in \Re^n$, $u_i \in \Re$ and $y_i \in \Re$ are the state, input and output of the $i$th agent, respectively; $\boldsymbol\theta_i \in \Re^{d_i}$ is a vector of unknown constants; $\boldsymbol\varphi_{i,k} \in \Re^{d_i}$ is a known smooth nonlinear function;\footnote{Arguments of some functions/variables will be omitted or replaced by $(\cdot)$ hereafter if no confusion is likely to occur.} $g_i(\boldsymbol x_i, t)$ is an unknown yet time-varying and state-dependent control coefficient, whose sign represents the control direction of the $i$th agent.

The control objective of this paper is to design a distributed control strategy such that:

\begin{itemize}
\item[${\cal O}_1$:]
All the internal signals are ensured to be SGUUB;
\end{itemize}

\begin{itemize}
\item[${\cal O}_2$:]
The output tracking error $e_i(t)$ for each agent is driven into an arbitrarily small neighborhood of the origin with an arbitrarily fast convergence rate.
\end{itemize}

To proceed, the following assumptions are in need.

\begin{assumption}\label{Assumption 1}
The augmented graph $\bar{\cal G}$ contains a spanning tree with the leader being the root node.
\end{assumption}

\begin{assumption}\label{Assumption 3}
The desired trajectory $y_0(t)$ and its time derivatives up to $n$th order are bounded and piecewise continuous in $t$.
\end{assumption}

\begin{assumption}\label{Assumption 2}
There exist two unknown constants $g_{i1}$ and $g_{i2}$ such that $g_i(\cdot) \in [g_{i1}, g_{i2}]$ with $g_{i1}g_{i2}>0$.
\end{assumption}

\begin{remark}\label{Remark 2}
\emph{Assumptions} \ref{Assumption 1}--\ref{Assumption 2} are reasonable. \emph{Assumption} \ref{Assumption 1} indicates that ${\boldsymbol{L}}+{\boldsymbol{B}}$ is a nonsingular ${\cal M}$-matrix with ${\boldsymbol{ B}}={\rm diag}(b_1,...,b_N)$ \cite{bechlioulis2016decentralized, katsoukis2021low}.
\emph{Assumption} \ref{Assumption 3} is widely used in the existing results for strict-feedback systems \cite{krstic1995nonlinear}.
\emph{Assumption} \ref{Assumption 2} essentially implies that the control direction of each agent is unknown, which, however, is sufficient to ensure the controllability of the system (see \cite{ge2003robust, huang2022distributed, bechlioulis2016decentralized} and the references therein).
\end{remark}

\begin{remark}\label{Remark 10}
It is noted that the unknown control coefficients in the system \eqref{eq.15} are allowed to be time-varying and even state-dependent, which is more general than the cases considered in \cite{chen2013adaptive, wang2020adaptive, liu2014adaptive, ding2015adaptive, psillakis2016consensus, wang2018cooperative}, \cite{huang2018fully, wang2020adaptiveb}, since the control coefficients in \cite{chen2013adaptive, wang2020adaptive, liu2014adaptive, ding2015adaptive, psillakis2016consensus, wang2018cooperative} are limited to be constant and the time-varying and state-dependent term in \cite{huang2018fully, wang2020adaptiveb} is required to be known. Clearly, the system models involved therein are essentially some special cases of ours.
In practice, there are numerous real practical systems, e.g., multi-robot systems, multi-wind power systems and swarm unmanned aerial systems \cite{huang2022distributed}, that can be boiled down to such systems. The reliability and safety of the whole system could be compromised when unknown non-identical control directions and mismatched uncertainties are involved, making the control design much more challenging. In order to overcome such obstacles, in this paper we develop an indirect performance guaranteed control framework consisting of distributed robust filters and Nussbaum-based backstepping adaptive controllers.
\end{remark}

\subsection{Some Useful Lemmas}
In this subsection, some necessary lemmas are introduced.
\begin{lemma}\label{Lemma 1}\cite{qu2009cooperative}
Consider a nonsingular ${\cal M}$-matrix ${\boldsymbol W} \in \Re^{N \times N}$. There exists a diagonal positive definite matrix $\boldsymbol P=(\rm{diag(\bar {\boldsymbol q})})^{-1}$ with $\bar {\boldsymbol q} = \boldsymbol W^{-1}\boldsymbol 1_N$, such that $\boldsymbol P \boldsymbol W+\boldsymbol W^{T}\boldsymbol P$ is also positive definite.
\end{lemma}

\begin{lemma}\label{Lemma 2}\cite[\emph{Theorem} 1]{verginis2019robust}
Let $\Omega \in \Re^n\times \Re_{\geq 0}$ be an open set. Consider a function $\boldsymbol f(\boldsymbol \zeta, t):\Omega \rightarrow \Re^n $ such that: (1) For every $\boldsymbol \zeta\in\Re^n$, the function $t\rightarrow \boldsymbol f(\boldsymbol \zeta, t)$ defined on $\Omega_t:=\{t:(\boldsymbol \zeta,t)\in\Omega\}$ is measurable. For every $t\in \Re_{\geq 0}$, the function $\boldsymbol \zeta\rightarrow \boldsymbol f(\boldsymbol \zeta, t)$ defined on $\Omega_\zeta:=\{\boldsymbol \zeta:(\boldsymbol \zeta,t)\in\Omega\}$ is continuous; (2) For every compact set $S \subset \Omega$, there exist constants $L_1$ and $L_2$ such that $\|\boldsymbol f(\boldsymbol \zeta, t)\|\leq L_1$ and $\|\boldsymbol f(\boldsymbol \zeta, t)- \boldsymbol f(\boldsymbol y, t)\| \leq L_2\|\boldsymbol \zeta-\boldsymbol y\|$, $\forall(\boldsymbol \zeta,t),(\boldsymbol y, t)\in S$. Then, the initial value problem $\dot{\boldsymbol \zeta} =\boldsymbol f(\boldsymbol \zeta, t)$ with $\boldsymbol \zeta_0= \boldsymbol \zeta(t_0)$, for some $(\boldsymbol \zeta_0, t_0)\in\Omega$, has a unique and maximal solution defined on $[t_0,T_{\mathrm{max}})$ with $T_{\mathrm{max}}>t_0$ such that $(\boldsymbol \zeta,t)\in \Omega$ for all $t\in [t_0,T_{\mathrm{max}})$.
\end{lemma}


\begin{lemma}\label{Lemma 3}\cite[\emph{Theorem} 2]{verginis2019robust}
Suppose that the conditions of \emph{Lemma} \ref{Lemma 2} hold in $\Omega$ and a maximal solution of the initial value problem $\dot{\boldsymbol \zeta} = \boldsymbol f(\boldsymbol \zeta, t), \boldsymbol \zeta_0=\boldsymbol \zeta(t_0)$ exists on $[t_0,T_{\mathrm{max}})$ such that $(\boldsymbol \zeta,t)\in \Omega,\forall t\in [0,T_{\mathrm{max}})$. Then, either $T_{\mathrm{max}}=\infty$ or $ \lim_{t\rightarrow T ^-_{\mathrm{max}}}\Big[\| {\boldsymbol \zeta}\|+\frac{1}{d_{S}(( {\boldsymbol \zeta},t),\partial \Omega)}\Big]=\infty$, where $d_{S}$ is the minimum distance from $({\boldsymbol \zeta},t)$ to the edge of $\Omega$, i.e., $\partial \Omega$.
\end{lemma}

Next, in order to deal with the time-varying coefficients, a type of special Nussbaum function is introduced.

\begin{definition}\cite[\emph{Definition} 4.2]{chen2019nussbaum}\label{Definition 1}
A continuously differentiable function $N(s) \in \Re$ is called a type of B-$K$ Nussbaum function, if, for a constant $K>1$, it satisfies
\begin{align}
\mathop {\lim }\limits_{\chi \to \infty } \frac{1}{\chi}{\int_0^\chi {{N^ + }(s)ds}} = \infty, ~\mathop {\lim }\limits_{\chi \to \infty } \sup \frac{{\int_0^\chi {{N^ + }(s)ds } }}{{\int_0^\chi {{N^ - }(s)ds } }} \ge K, \nonumber\\
\mathop {\lim }\limits_{\chi \to \infty } \frac{1}{\chi}{\int_0^\chi {{N^ - }(s)ds}} = \infty, ~\mathop {\lim }\limits_{\chi \to \infty } \sup \frac{{\int_0^\chi {{N^ - }(s)ds } }}{{\int_0^\chi {{N^ + }(s )ds } }} \ge K, \nonumber
\end{align}
where $N^+(s)$ and $N^-(s)$ denote its positive and negative truncated functions, that is
\begin{align}\label{eq.44}
N^+(s)={\rm max} \left\{ 0, N(s) \right\}, ~N^-(s)={\rm max} \left\{ 0, -N(s) \right\}.
\end{align}
\end{definition}

Then the following lemma regarding Nussbaum function is introduced, which is crucial for the control design and stability analysis as seen later.

\begin{lemma}\label{Lemma 4}
Consider two continuously differentiable functions $V(t): [0, \infty)\mapsto \Re^+$ and $\chi(t): [0, \infty)\mapsto \Re$. Let $\eta(t): [0, \infty)\mapsto [\eta_1, \eta_2]$ with $\eta_1\eta_2>0$. If
\begin{align}\label{eq.17}
\dot{V}(t)\leq (\eta(t)N(\chi(t))+a)\dot{\chi}(t), \quad \dot{\chi}(t) \in \Re
\end{align}
for any constant $a$ and any Nussbaum function (type B-$K$) $N(\chi)$ with $K>{\rm max}\left\{ \eta_2/\eta_1, \eta_1/\eta_2 \right\}$, then $V(t)$ and $\chi(t)$ are bounded for all $t \in [0, \infty)$.
\end{lemma}

\textbf{\emph{Proof}}: See the Appendix. \hfill $\blacksquare$

\begin{remark}\label{Remark 4}
It should be emphasized that \emph{Lemma} \ref{Lemma 4} plays a vital role in our work since it broadens the applicability of the Nussbaum function related stability theory to a more general case. More precisely, it relaxes the condition imposed on $\dot{\chi}(t)$ by extending its domain to $\Re$, rendering the result developed in \cite{chen2019nussbaum} as a special case because $\dot{\chi}(t)>0$ is required therein.
Another salient merit hiding in \emph{Lemma} \ref{Lemma 4} is that its results hold for all $t \geq 0$ instead of only during a finite time interval $[0,t_f)$ as that in \cite{xudong1998adaptive, ge2003robust}, which significantly simplifies the complexity during the stability analysis since no extra and tedious process is needed to extend $t_f$ to infinity.
\end{remark}

\section{Distributed Control Design with Indirect Performance-guaranteed Framework}
In this section, we will divide the the control design into two parts that provide an effective solution to the underlying problem. The first part shows the design and analysis of distributed filter to estimate the reference trajectory with guaranteed-performance. The second part presents the backstepping design procedure of an adaptive control scheme to achieve asymptotically tracking with prescribed transient performance. For convenience, a block diagram is provided in Fig. \ref{Fig.6} to illustrate the internal relationship between them.

\subsection{Filter Design and Analysis}
Since only part of the followers have access to the desired trajectory, a distributed robust filter is constructed for each agent to estimate such trajectory with prescribed performance.
\subsubsection{Filter Design}
For the $i$th $~(i=1,...,N)$ agent, the filter is designed as
\begin{align}\label{eq.1}
\hat{y}^{(n)}_i=\nu_i
\end{align}
where $\hat{y}_i$ is the estimation of $y_0$; $\hat{y}_i, \dot{\hat{y}}_i,...,\hat{y}^{(n-1)}_i$ are the states of the filter, and $\nu_i$ is the input that will be given later. Let
\begin{align}\label{eq.3}
\bar{\delta}_i &=\left(\frac{d}{dt}+\lambda\right)^{n-1}\delta_i=\sum\limits_{{{k}} = 0}^{n - 1} C_{n - 1}^k \lambda^k \delta^{(n-1-k)}_i, \\
\label{eq.53}
z_i &=\sum\limits_{j \in {{\cal N}_i}} {{a_{ij}}(\hat{y}_i-\hat{y}_j)}  + {b_i}(\hat{y}_i - {y_0}), \\
\label{eq.4}
\bar{z}_i &=\left(\frac{d}{dt}+\lambda\right)^{n-1}z_i=\sum\limits_{{{k}} = 0}^{n - 1} C_{n - 1}^k \lambda^k z^{(n-1-k)}_i,
\end{align}
where $\delta_i$ is given in \eqref{eq.2} and $\lambda$ is a positive constant.
It can be seen from \eqref{eq.53} and \eqref{eq.4} that both $z_i$ and $\bar{z}_i$ only involve local information related to $i$th agent and its neighbors, and thus can be used for distributed controller design, while $\delta_i$ and $\bar{\delta}_i$ can be only used for stability analysis.


To establish a quantitative relationship between the filter error and the consensus error via a compact form, we define $\boldsymbol z = {[{{z}_1},...,{{z}_N}]^T} \in \Re^N$, $\bar {\boldsymbol z} = {[{{\bar z}_1},...,{{\bar z}_N}]^T} \in \Re^N$,
$\boldsymbol \delta = {[{{\delta}_1},...,{{\delta}_N}]^T} \in \Re^N$ and $\bar {\boldsymbol\delta} = {[{{\bar \delta}_1},...,{{\bar \delta}_N}]^T} \in \Re^N$,
then
\begin{align}\label{eq.5}
\boldsymbol z =({\boldsymbol{ L}}+{\boldsymbol{ B}})\boldsymbol \delta,
\quad\bar {\boldsymbol z} =({\boldsymbol{ L}}+{\boldsymbol{ B}})\bar {\boldsymbol\delta}.
\end{align}
Owing to \emph{Assumption} \ref{Assumption 1}, it follows from \eqref{eq.5} that
\begin{align}\label{eq.55}
\left\| \boldsymbol\delta  \right\| \le \frac{{\left\| \boldsymbol z \right\|}}{{{\sigma _{\min }}(\boldsymbol L + \boldsymbol B)}},
\end{align}
which indicates that the prescribed performance imposed on $\|\boldsymbol\delta\|$ (or $\delta_i$) can be boiled down to that on $\|\boldsymbol z\|$ (or $z_i$). Recalling the definition of ${\bar z}_i$ in \eqref{eq.4}, we have $z_i(p)=\bar z_i(p)/(p+\lambda)^{n-1}$, where $p$ denotes the Laplace operator. Since $1/(p+\lambda)^{n-1}$ is stable, it is clear that the performance bounds on ${\bar z}_i$ can be directly translated into that on $z_i$ \cite{song2016adaptive}. Hence, the underlying problem will be solved if the prescribed performance bounds of ${\bar z}_i$ are ensured. Consequently, the relationship between $\bar{z}_i$ and $\delta_i$ in terms of performance metrics is summarized in the following proposition.

\begin{proposition}\label{Proposition 1}
Suppose that $\bar z_i(t)$ evolves within the following performance bounds:
\begin{align}\label{eq.49}
|\bar{z}_i(t)|<\rho_{i}(t),
\end{align}
where $\rho_{i}(t)=(\rho_{0}-\rho_{\infty})e^{-\iota t}+\rho_{\infty}$
is a performance function with $\lambda>\iota$, then it holds that:
\end{proposition}

\begin{itemize}
\item[$(1)$]
The converge rate of $\boldsymbol\delta$ is faster than ${\rm exp}(-\iota t)$;
\end{itemize}

\begin{itemize}
\item[$(2)$]
The size of the steady-state error is smaller than $\sqrt{N}\rho_{\infty}/(\lambda^{n-1}\sigma_{\rm min}(\boldsymbol{ L}+\boldsymbol{ B}))$.
\end{itemize}

\textbf{\emph{Proof}}: See the Appendix. \hfill $\blacksquare$

\begin{remark}\label{Remark 11}
\emph{Proposition} 1 shows that based on \eqref{eq.4} and \eqref{eq.55}, the performance-guaranteed problem for the filter error  $\delta_i(t)$ regarding convergence rate and steady-state error can be recast into the one for $\bar z_i(t)$ through a quantitative way. In this sense, the convergence rate of $\delta_i(t)$ can be made arbitrarily fast by selecting $\iota$ large enough and the steady-state error can be made arbitrarily small by choosing $\rho_{\infty}$ small enough. In \eqref{eq.55}, the calculation of $\sigma_{\rm min}(\boldsymbol{ L}+\boldsymbol{ B})$ requires the global topology information, which can be avoided by computing its lower bound $\kappa= ((N-1)/N)^{(N-1)/2}/(N^2+N-1)$ \cite{hong1992lower}, and thus contributing to a completely distributed scheme.
\end{remark}

In the remainder of this subsection, we focus on the achievement of the guaranteed performance for $\bar z_i$. To this end, the dynamics of $\bar z_i$ should be first deduced.

Taking the time derivative of $\bar \delta_i$ along \eqref{eq.2}, \eqref{eq.1} and \eqref{eq.3} yields
\begin{align}
\dot{\bar{\delta}}_i&=\delta^{(n)}_i+\sum\limits_{k = 1}^{n - 1}\epsilon_k \delta^{(n-k)}_i \nonumber\\
&=\nu_i-y^{(n)}_0+\sum\limits_{k = 1}^{n - 1} \epsilon_k \left( {\hat{y}^{(n-k)}_i - y_0^{(n-k)}} \right), \nonumber
\end{align}
where $\epsilon_k=C_{n - 1}^k\lambda^k$. By adding and subtracting $b_i\left( y^{(n)}_0+\sum\nolimits_{k = 1}^{n - 1} \epsilon_k y^{(n-k)}_0 \right)$, one obtains that
\begin{align}\label{eq.56}
\dot{\bar{\delta}}_i=&\nu_i+\underbrace {\sum\limits_{k = 1}^{n - 1} \epsilon_k \hat{y}^{(n-k)}_i-b_iy^{(n)}_0-b_i\sum\limits_{k = 1}^{n - 1} \epsilon_k {y_0^{(n-k)}}}_{=\varpi_i} \nonumber\\
&+\underbrace{(b_i-1)\left( y^{(n)}_0+\sum\limits_{k = 1}^{n - 1} \epsilon_k y^{(n-k)}_0 \right)}_{=\Delta_i}
\end{align}
where $\varpi_i$ is a bounded term that will not be used in control design; $\Delta_i$ is an unknown yet bounded function from \emph{Assumption} \ref{Assumption 3}.

Let $\boldsymbol\nu=\left[ \nu_1, ... , \nu_N \right]^T$, $\boldsymbol\varpi=\left[ \varpi_1, ... , \varpi_N \right]^T$ and $\boldsymbol\Delta=\left[ \Delta_1, ... , \Delta_N \right]^T$, then \eqref{eq.56} can be rewritten as
\begin{align}\label{eq.57}
\dot{\bar{\boldsymbol {\delta}}}=\boldsymbol\nu+\boldsymbol\varpi+\boldsymbol\Delta.
\end{align}
From \eqref{eq.5} and \eqref{eq.57}, it is derived that
\begin{align}\label{eq.6}
\dot{\bar{\boldsymbol z}}=({\boldsymbol{L}}+{\boldsymbol{B}})\dot{\bar{\boldsymbol {\delta}}}=({\boldsymbol{ L}}+{\boldsymbol{ B}})(\boldsymbol \nu+\boldsymbol\varpi+\boldsymbol\Delta).
\end{align}
To achieve the prescribed performance for $\bar{z}_i$, we define the following variables
\begin{align}\label{eq.50}
\zeta_{\bar{z}_i}(t)&=\frac{\bar{z}_i}{\rho_{i}},\\
\label{eq.51}
O_i(\zeta_{\bar{z}_i})&={\rm ln}\left(\frac{1+\zeta_{\bar{z}_i}}{1-\zeta_{\bar{z}_i}}\right),\\
\label{eq.59}
M_i(\zeta_{\bar{z}_i})&=\frac{1}{1-\zeta^2_{\bar{z}_i}},
\end{align}
where $\rho_{i}(t)$ is the performance function of $\bar{z}_i(t)$. It can be checked that $O_i: (-1,1) \rightarrow (-\infty, +\infty)$ is a smooth and strictly increasing function exhibiting the following properties:

\noindent 1) $\lim _{\zeta_{\bar{z}_i} \to -1 } O_i = -\infty$;

\noindent 2) $\lim _{\zeta_{\bar{z}_i} \to +1 } O_i = +\infty$;

\noindent 3) $O_i(0)=0$.

\noindent In other words, if $O_i \in L_\infty$, it holds that $|\zeta_{\bar{z}_i}|<1$, which indicates that the performance imposed on $\bar{z}_i$ is guaranteed.
To this end, the distributed control for filter \eqref{eq.1} is designed as
\begin{align}\label{eq.7}
\nu_i(\zeta_{\bar{z}_i},t)=-\frac{c_i}{\rho_{i}}\cdot\frac{O_i}{M_i}
\end{align}
where $c_i$ is a positive constant, and $\nu_i$ is well defined since $M_i$ is nonsingular.

\subsubsection{Stability Analysis}
The main result of the distributed performance guaranteed filter design can be stated in the following theorem.
\begin{theorem}\label{Theorem 1}
Consider the closed-loop system consisting of $N$ filters \eqref{eq.1} with the input \eqref{eq.7} under \emph{Assumptions} \ref{Assumption 1}--\ref{Assumption 3}. For an arbitrarily small positive constant $\varrho$, there exists a ball $\mathcal{B}_{r_0}=\{\zeta_{\bar{z}_i} | |\zeta_{\bar{z}_i}|<r_0=1-\varrho\}$ such that $\zeta_{\bar{z}_i}(0) \in \mathcal{B}_{r_0}$,
then all the internal signals are SGUUB and the prescribed performance of $\bar z_i(t)$ is ensured.
\end{theorem}

\textbf{\emph{Proof}}:
In order to perform such proof in a compact matrix form, we define the normalized error vector along \eqref{eq.50} as
\begin{align}
\boldsymbol \zeta=\left[ \zeta_{\bar{z}_1}, ... , \zeta_{\bar{z}_N} \right]^T \buildrel \Delta \over = \boldsymbol H^{-1}\bar{\boldsymbol z}
\end{align}
in which $\boldsymbol H={\rm diag}(\rho_{1}(t), ... , \rho_{N}(t))$. Differentiating $\boldsymbol \zeta$ w.r.t. $t$ obtains
\begin{align}\label{eq.8}
\dot{\boldsymbol \zeta}=\boldsymbol H^{-1}( \dot{\bar{\boldsymbol z}}-\dot{\boldsymbol H}\boldsymbol \zeta).
\end{align}
Using \eqref{eq.6} and \eqref{eq.7}, \eqref{eq.8} can be derived as
\begin{align}\label{eq.9}
\dot{\boldsymbol\zeta}=\boldsymbol f(\boldsymbol\zeta, t)=&\boldsymbol H^{-1}\left( ({\boldsymbol{ L}}+{\boldsymbol{ B}}) \left( -\overline{\boldsymbol C} \boldsymbol H^{-1}\boldsymbol M\boldsymbol O
\right. \right. \nonumber\\
&\left. \left. +\boldsymbol\varpi+\boldsymbol\Delta \right)-\dot{\boldsymbol H}\boldsymbol\zeta \right),
\end{align}
where $\overline{\boldsymbol C}={\rm diag}(c_1, ... , c_N)$ and
\begin{align}\label{eq.10}
\boldsymbol M={\rm diag}\left( M_1, ... , M_N \right), ~
\boldsymbol O=\left[ O_1, ... , O_N \right]^T
\end{align}
with $M_i$ and $O_i$ defined as in \eqref{eq.51} and \eqref{eq.59}, respectively. Define an open ball $\mathcal{B}_{r_1}$ as
\begin{align}
\mathcal{B}_{r_1}=\{\boldsymbol\zeta | |\zeta_{\bar{z}_i}|<r_1=1\}, \quad i=1,...,N.
\end{align}
Then $\dot{\boldsymbol\zeta}$-dynamics in \eqref{eq.9} is well-defined for all $\boldsymbol\zeta \in \mathcal{B}_{r_1}$. To go on with the proof, we need to consider each of the following three phases. First, we address the existence and uniqueness of a maximal solution $\boldsymbol\zeta(t)$ of \eqref{eq.9} over the ball $\mathcal{B}_{r_1}$ during a time interval $[0, T_{\mathrm{max}})$. Second, we prove that all internal signals in \eqref{eq.9} are bounded and $\boldsymbol\zeta(t)$ remains strictly within a compact subset of $\mathcal{B}_{r_1}$ for $t \in [0, T_{\mathrm{max}})$ with the proposed control scheme \eqref{eq.7}. Third, we show that $T_{\mathrm{max}}$ can be extended to $+\infty$ and subsequently $\left| {{{\bar z}_i}} \right| < \rho_{i}(t) ~(i=1, ... , N)$ is assured for $\forall t \geq 0$.


\textbf{\emph{Phase 1: Existence and uniqueness of a feasible solution.}} Because $\rho_{i}(t)$ has been selected to satisfy $\zeta_{\bar{z}_i}(0) \in \mathcal{B}_{r_0} ~(i=1, ... , N)$, it holds that $\boldsymbol\zeta(0) \in \mathcal{B}_{r_1}$. Moreover, owing to the fact that the performance function $\rho_{i}(t)$ and the desired signal $y_0(t)$ are bounded and continuously differentiable functions w.r.t. $t$, it is concluded that nonlinear function $\boldsymbol f(\boldsymbol\zeta, t)$ is locally Lipschitz in $\boldsymbol\zeta$, uniformly in $t$. Hence, the conditions of \emph{Lemma} \ref{Lemma 2} are sufficed, and the initial value problem in \eqref{eq.9} has a unique and maximal solution over the time interval $[0, T_{\mathrm{max}})$, such that $\boldsymbol\zeta(t) \in \mathcal{B}_{r_1}, \forall t \in [0, T_{\mathrm{max}})$, that is
\begin{align}\label{eq.11}
\zeta_{\bar{z}_i} \in (-1, 1), \quad i=1, ... , N , \quad\forall t \in [0, T_{\mathrm{max}}).
\end{align}

\textbf{\emph{Phase 2:  Boundedness of all the internal signals over $[0, T_{\mathrm{max}})$.}} Since $\overline{\boldsymbol C}$ is a diagonal with positive entries, $({\boldsymbol{L}}+{\boldsymbol{ B}})\overline{\boldsymbol C}$ is a nonsingular ${\cal M}$-matrix under \emph{Assumption} \ref{Assumption 1} \cite{bechlioulis2016decentralized, katsoukis2021low}. Therefore, according to \emph{Lemma} \ref{Lemma 1}, there exists a diagonal positive definite matrix $\boldsymbol P$ defined as $\boldsymbol P={\left( {\rm diag}(\boldsymbol q) \right)^{ - 1}}$ with $\boldsymbol q=((\boldsymbol{L}+\boldsymbol{B})\overline{\boldsymbol C})^{-1}\boldsymbol 1_N$. Then take a Lyapunov function candidate as
\begin{align}
V_O=\frac{1}{2}\boldsymbol O^T\boldsymbol P\boldsymbol O, \nonumber
\end{align}
where $\boldsymbol O$ is given in \eqref{eq.10} that is also well-defined over $\mathcal{B}_{r_1}$. The time derivative of $\boldsymbol O$ is derived along \eqref{eq.9} as
\begin{align}
\dot{\boldsymbol O}=&2\boldsymbol M\boldsymbol H^{-1} \left( (\boldsymbol{L} + \boldsymbol{ B} )\left(-\overline{\boldsymbol C}\boldsymbol H^{-1}\boldsymbol M\boldsymbol O +\boldsymbol\varpi+\boldsymbol\Delta \right.)-\dot{\boldsymbol H}\boldsymbol\zeta \right), \nonumber
\end{align}
which leads to
\begin{align}\label{eq.12}
\dot{V}_O=&-2\boldsymbol O^T\boldsymbol P\boldsymbol M\boldsymbol H^{-1}(\boldsymbol{L}+\boldsymbol{ B})\overline{\boldsymbol C}\boldsymbol H^{-1}\boldsymbol M\boldsymbol O \nonumber\\
&+2\boldsymbol O^T\boldsymbol P\boldsymbol M\boldsymbol H^{-1}\left((\boldsymbol{L}+\boldsymbol{ B})(\boldsymbol\varpi+\boldsymbol\Delta)-\dot {\boldsymbol H}\boldsymbol\zeta\right)  \nonumber\\
=&-\boldsymbol O^T\boldsymbol M\boldsymbol H^{-1}\underbrace{\left( \boldsymbol P(\boldsymbol{L}+\boldsymbol{ B})\overline{\boldsymbol C}+\overline{\boldsymbol C}(\boldsymbol{L}+\boldsymbol{B})^T\boldsymbol P \right)}_{\boldsymbol{Q}} \nonumber\\
&\times \boldsymbol H^{-1}\boldsymbol M\boldsymbol O + 2\boldsymbol O^T\boldsymbol M\boldsymbol H^{-1} \nonumber\\
&\times \boldsymbol P\left((\boldsymbol{L}+\boldsymbol{B})(\boldsymbol\varpi+\boldsymbol\Delta)-\dot {\boldsymbol H}\boldsymbol\zeta\right).
\end{align}
Note that $\boldsymbol P$ and $(\boldsymbol L+\boldsymbol B)$ are constant matrices, $\boldsymbol\zeta$ and $\boldsymbol\varpi$ are bounded for $t \in [0, T_{\mathrm{max}})$ under \eqref{eq.11}, $\boldsymbol\Delta$ and $\dot {\boldsymbol H}$ are bounded due to the boundedness of $\dot{\rho}_{i}$ and $y^{(k)}_0 ~(k=0, ... , n)$ by construction and assumption. Then there exists a positive constant $\overline{L}$ such that
\begin{align}
\left\| {\boldsymbol P\left( {\left( {\boldsymbol{L} + \boldsymbol{B}} \right) (\boldsymbol\varpi+\boldsymbol\Delta)- \dot{\boldsymbol H}\boldsymbol\zeta} \right)} \right\| \leq \overline{L} \nonumber
\end{align}
for $\forall t \in [0, T_{\mathrm{max}})$. Due to the fact that $\lambda_{\mathrm{min}}(\boldsymbol M) \geq 1$, $\lambda_{\mathrm{min}}(\boldsymbol H^{-1}) \geq \lambda_{\boldsymbol H^{-1}}=1/
{\mathop {\max }\limits_{i \in 1,...,N} \{ {\rho_{i}(0)}\} }$
and the matrix $\boldsymbol Q$ is positive definite, then $\dot{V}_O$ in \eqref{eq.12} can be bounded as
\begin{align}\label{eq.13}
\dot{V}_O \leq &-\lambda_{\mathrm{min}}(\boldsymbol Q)\left\|\boldsymbol O^T\boldsymbol M\boldsymbol H^{-1} \right\|^2 + 2\left\|\boldsymbol O^T\boldsymbol M\boldsymbol H^{-1} \right\|\overline{L}  \nonumber\\
\leq &-\frac{1}{2}\lambda_{\mathrm{min}}(\boldsymbol Q)\lambda^2_{\boldsymbol H^{-1}}\left\|\boldsymbol O \right\|^2 + \frac{2\overline{L}^2}{\lambda_{\mathrm{min}}(\boldsymbol Q)},
\end{align}
where the fact that
\begin{align}
\left\|\boldsymbol O^T\boldsymbol M\boldsymbol H^{-1} \right\|\overline{L} \leq & \frac{\lambda_{\mathrm{min}}(\boldsymbol Q)}{4}\left\|\boldsymbol O^T\boldsymbol M\boldsymbol H^{-1} \right\|^2 + \frac{4\overline{L}^2}{\lambda_{\mathrm{min}}(\boldsymbol Q)} \nonumber
\end{align}
is applied owing to Young's inequality. Furthermore, it can be concluded from \eqref{eq.13} that
\begin{align}
\left\|\boldsymbol O(\boldsymbol\zeta) \right\| \leq \bar O = \max \left\{ {\left\| {\boldsymbol O\left( {\boldsymbol\zeta(0)} \right)} \right\|,\frac{{2 \overline{L}\sqrt {\frac{{{\lambda _{\mathrm{max} }}(\boldsymbol Q)}}{{{\lambda _{\mathrm{min} }}(\boldsymbol Q)}}} }}{{{\lambda _{\min }}(\boldsymbol Q){\lambda _{{\boldsymbol H^{ - 1}}}}}}} \right\} \nonumber
\end{align}
for $\forall t \in [0, T_{\mathrm{max}})$, which, combining \eqref{eq.10} with \eqref{eq.51}, leads to
\begin{align}\label{eq.14}
-1 < \frac{e^{-\bar{O}}-1}{e^{-\bar{O}}+1} = \underline{\zeta}_{\bar{z}_i} \leq \zeta_{{\bar{z}_i}} \leq \overline{\zeta}_{\bar{z}_i} = \frac{e^{\bar{O}}-1}{e^{\bar{O}}+1} < 1
\end{align}
for $i=1, ... , N$ and $t \in [0, T_{\mathrm{max}})$, then the boundedness of $O_i$ and $M_i$ is ensured from \eqref{eq.51} and \eqref{eq.59}. Therefore, the control signals $\nu_i$ in \eqref{eq.7} are bounded for $\forall t \in [0, T_{\mathrm{max}})$.

\textbf{\emph{Phase 3:  Extension to $T_{\mathrm{max}}=+\infty$.}} In Phase 1, it has been shown that the closed-loop system \eqref{eq.9} has a unique and maximal solution $\boldsymbol\zeta(t)$ over $t \in [0, T_{\mathrm{max}})$. By \emph{Lemma} \ref{Lemma 3}, either $T_{\mathrm{max}}=\infty$ or $\lim_{t\rightarrow T ^-_{\mathrm{max}}}\Big[\| {\boldsymbol\zeta}\|+\frac{1}{d_{S}(( {\boldsymbol\zeta},t),\partial \mathcal{B}_{r_1}(\boldsymbol\zeta))}\Big]=\infty$. Moreover, notice by \eqref{eq.14} that for all $t \in [0, T_{\mathrm{max}})$
\begin{align}
\boldsymbol\zeta(t) &\in \mathcal{B}'_{r_1} \buildrel \Delta \over = \{\boldsymbol\zeta | \underline{\zeta}_{\bar{z}_i}<\zeta_{\bar{z}_i}<\overline{\zeta}_{\bar{z}_i},i=1,...,N\}
\end{align}
with $\mathcal{B}'_{r_1} \subset \mathcal{B}_{r_1}$.
Hence, it can be deduced that $\| {\boldsymbol\zeta}\|+\frac{1}{d_{S}(( {\boldsymbol\zeta},t),\partial \mathcal{B}_{r_1}(\boldsymbol\zeta))}<\infty$, ~$\forall t \in [0, T_{\mathrm{max}})$. Therefore, it holds that $T_{\mathrm{max}}=\infty$. Consequently, all internal signals remain bounded and $-1 < \underline{\zeta}_{\bar{z}_i} \leq \zeta_{{\bar{z}_i}} \leq \overline{\zeta}_{\bar{z}_i} < 1 ~(i=1, ... , N)$ for $\forall t \geq 0$. Finally, it is concluded from \eqref{eq.50} and \eqref{eq.14} that
\begin{align}\label{eq.43}
-\rho_{i}(t) < \underline{\zeta}_{\bar{z}_i}\rho_{i}(t) \leq {\bar{z}_i} \leq \overline{\zeta}_{\bar{z}_i}\rho_{i}(t) < \rho_{i}(t)
\end{align}
for $\forall t \geq 0$. Then the synchronization control problem with prescribed performance is achieved. The proof is
completed.
\hfill $\blacksquare$

\begin{remark}\label{Remark 6}
It should be noted that the developed distributed filter \eqref{eq.7} is indispensable in the framework of our scheme. Firstly and most importantly, it is able to reconstruct the desired trajectory for each agent (in a performance guaranteed manner) and thus decouple the unknown non-identical control direction problem among them, which, at the same time, simplifies the controller design, i.e., implementation phase, for each agent since only a set of identical (instead of non-identical as compared with the method in \cite{wang2020adaptive, wang2020adaptiveb}) Nussbaum functions are utilized therein. Secondly, it avoids excessively increasing structural complexity and aggravating the computational burden of the whole scheme since the developed filter remains at a lower complexity as compared to those in \cite{cai2015leader, huang2017smooth, chen2015observer, liu2017adaptive} by circumventing the online updating of any adaptive terms. Thirdly, the filter error $\delta_i$ is steered into an arbitrarily pre-assigned residual set with  an arbitrarily pre-defined convergence rate, which is different from \cite{wang2020adaptiveb, huang2017smooth, feng2016distributed}, since the transient performance/steady state behavior therein cannot be quantificationally guaranteed.
\end{remark}

\subsection{Adaptive Backstepping Design and Analysis}
In this subsection, an adaptive backstepping tracking control algorithm is presented for each agent with prescribed performance in the presence of unknown non-identical control directions and time-varying coefficients.
\subsubsection{Controller Design}
\allowdisplaybreaks
We aim to design $u_i$ for each agent such that the output $y_i(t)$ asymptotically tracks the filtered variable $\hat{y}_i(t)$ with the corresponding tracking error $\varepsilon_i(t)$ preserved within specified performance bounds all the time, that is
\begin{align}\label{eq.19}
|\varepsilon_i(t)|<\beta_{i2}(t), \quad \forall t \geq 0
\end{align}
where $\beta_{i2}(t)$ is the performance function of $\varepsilon_i(t)$ as defined in \eqref{eq.52}. Similar to \eqref{eq.50} and \eqref{eq.51}, some auxiliary variables is introduced as
\begin{align}\label{eq.16}
\zeta_{\varepsilon_i}(t)&=\frac{\varepsilon_i}{\beta_{i2}},\\
\label{eq.20}
\xi_i(\zeta_{\varepsilon_i})&={\rm ln}\left(\frac{1+\zeta_{\varepsilon_i}}{1-\zeta_{\varepsilon_i}}\right).
\end{align}
As discussed in Section III-A, the performance bound $|\varepsilon_i(t)|<\beta_{i2}(t)$ is achieved as long as $\xi_i$ is bounded. Next, by following the standard backstepping design procedure \cite{krstic1995nonlinear}, we carry out the control design step by step. Let
\begin{align}
\label{eq.48}
e_{i,1}&=\xi_i, \\
\label{eq.21}
e_{i,k}&=x_{i,k}-\alpha_{i,k-1}-\hat{y}^{(k-1)}_i, \quad k=2,...,n
\end{align}
where $\alpha_{i,k-1}$ is the virtual controller to be designed.

\textbf{\emph{Step 1}}: From \eqref{eq.2}, \eqref{eq.15}, \eqref{eq.16}--\eqref{eq.48}, the time derivative of $e_{i,1}$ can be calculated as
\begin{align}\label{eq.24}
\dot{e}_{i,1}=\mu_i\left( \alpha_{i,1}+e_{i,2}+\boldsymbol\theta^T_i\boldsymbol\varphi_{i,1}-\zeta_{\varepsilon_i}\dot{\beta}_{i2} \right),
\end{align}
where $\mu_i=2/(\beta_{i2}(1-\zeta^2_{\varepsilon_i}))$ is well defined and strictly positive.
To stabilize \eqref{eq.24}, the virtual control $\alpha_{i,1}$ is designed as
\begin{align}\label{eq.23}
\alpha_{i,1}=-\frac{c_{i,1}e_{i,1}}{\mu_i}-\hat{\boldsymbol\theta}^T_i\boldsymbol\varphi_{i,1}+\zeta_{\varepsilon_i}\dot{\beta}_{i2},
\end{align}
where $c_{i,1}$ is a positive design parameter and $\hat{\boldsymbol\theta}_i$ is the estimate of $\boldsymbol\theta_i$. Substituting \eqref{eq.23} into \eqref{eq.24} gives
\begin{align}
\dot{e}_{i,1}=-c_{i,1}e_{i,1}+\mu_ie_{i,2}+\mu_i\tilde{\boldsymbol\theta}^T_i\boldsymbol\varphi_{i,1}, \nonumber
\end{align}
with $\tilde{\boldsymbol\theta}_i=\boldsymbol\theta_i-\hat{\boldsymbol\theta}_i$ being the estimate error. Choose a Lyapunov function candidate as
\begin{align}\label{eq.40}
V_{i,1}=\frac{1}{2}e^2_{i,1}+\frac{1}{2}\tilde{\boldsymbol\theta}^T_i\boldsymbol\Gamma^{-1}_i\tilde{\boldsymbol\theta}_i,
\end{align}
where $\boldsymbol\Gamma_i \in \Re^{d_i \times d_i}$ is a positive definite matrix. Inspired by \cite{krstic1995nonlinear}, we define the first tuning function as
\begin{align}
\boldsymbol\tau_{i,1}=\mu_ie_{i,1}\boldsymbol\varphi_{i,1}, \nonumber
\end{align}
then the time derivative of $V_{i,1}$ can be derived as
\begin{align}\label{eq.25}
\dot{V}_{i,1}=-c_{i,1}e_{i,1}^2+\mu_ie_{i,1}e_{i,2}+\tilde{\boldsymbol\theta}^T_i\boldsymbol\Gamma^{-1}_i
\left(\boldsymbol\Gamma_i\boldsymbol\tau_{i,1}-
\dot{\hat{\boldsymbol{\theta}}}_i\right),
\end{align}
where the term $\mu_ie_{i,1}e_{i,2}$ will be cancelled in the next step.

\begin{table}
\centering
\caption{Design of Step $k ~(k=2,...,n)$}
\begin{tabular}{|l|c|}
\hline
\tabincell{l}{Step 2:\\$\alpha_{i,2}=-c_{i,2}e_{i,2}-\mu_ie_{i,1}-\hat{\boldsymbol\theta}^T_i\boldsymbol\omega_{i,2}+
\frac{\partial{\alpha}_{i,1}}{\partial{x}_{i,1}}
x_{i,2}$\\$\quad\quad+\frac{\partial{\alpha}_{i,1}}{\partial\hat{y}_i}\dot{\hat{y}}_i+\sum\limits_{k = 0}^1 \frac{\partial{\alpha}_{i,1}}{\partial{\beta}^{(k)}_{i2}}\beta^{(k+1)}_{i2}+\frac{\partial{\alpha}_{i,1}}
{\partial{\hat{\boldsymbol\theta}}_{i}}\boldsymbol\Gamma_i\boldsymbol\tau_{i,2}$\\with\\$\boldsymbol\omega_{i,2}
=\boldsymbol\varphi_{i,2}-
\frac{\partial{\alpha}_{i,1}}{\partial{x}_{i,1}}\boldsymbol\varphi_{i,1}$\\$\boldsymbol\tau_{i,2}
=\boldsymbol\tau_{i,1}
+\boldsymbol\omega_{i,2}e_{i,2}$} & (T1.1)\\
\hline
\tabincell{l}{Step $k~(k=3,...,n-1)$:\\$\alpha_{i,k}=-c_{i,k}e_{i,k}-e_{i,k-1}-\hat{\boldsymbol\theta}^T_i\boldsymbol\omega_{i,k}
$ \\ $\quad\quad +\sum\limits_{j = 1}^{k - 1}\frac{\partial{\alpha}_{i,k-1}}{\partial{x}_{i,j}}\beta _{i2}^{(k + 1)} +\sum\limits_{j = 0}^{k - 1} \frac{{\partial {\alpha _{i,k - 1}}}}{{\partial \beta _{i2}^{(j)}}}\beta _{i2}^{(j + 1)}$ \\$\quad\quad +\sum\limits_{j = 1}^{k - 1}\frac{\partial{\alpha}_{i,k-1}}{\partial{x}_{i,j}}\beta _{i2}^{(k + 1)}+\frac{\partial{\alpha}_{i,k-1}}{\partial{\hat{\boldsymbol\theta}}_
{i}}\boldsymbol\Gamma_i\boldsymbol\tau_{i,k}$\\$\quad\quad+\sum\limits_{j = 2}^{k - 1}\frac{\partial{\alpha}_{i,j-1}}{\partial{\hat{\boldsymbol\theta}}_
{i}}\boldsymbol\Gamma_i\boldsymbol\omega_{i,k}e_{i,j}$\\with\\$\boldsymbol\omega_{i,k}=\boldsymbol\varphi_{i,k}
-\sum\nolimits_{j = 1}^{k - 1}
\frac{\partial{\alpha}_{i,k-1}}{\partial{x}_{i,j}}\boldsymbol\varphi_{i,j}$\\$\boldsymbol\tau_{i,k}
=\boldsymbol\tau_{i,k-1}
+\boldsymbol\omega_{i,k}e_{i,k}$} & (T1.2)
\\ \hline
\tabincell{l}{Step $n$:\\$u_i=N_i(\chi_i)\bar{u}_i$\\
$\bar{u}_i=c_{i,n}e_{i,n}+e_{i,n-1}+\hat{\boldsymbol\theta}^T_i\boldsymbol\omega_{i,n}-\nu_i$\\
$\quad -\sum\limits_{k = 1}^{n - 1}\frac{\partial{\alpha}_{i,n-1}}{\partial{x}_{i,k}}x_{i,k+1}
-\sum\limits_{k = 0}^{n - 1} \frac{{\partial {\alpha _{i,n - 1}}}}{{\partial \beta _{i2}^{(k)}}}\beta _{i2}^{(k + 1)}$\\
$\quad -\sum\limits_{k = 1}^{n - 1} {\frac{{\partial {\alpha _{i,n - 1}}}}{{\partial \hat{y}^{(k-1)}_i}}\hat{y}^{(k)}_i}-\frac{\partial{\alpha}_{i,n-1}}{\partial{\hat{\boldsymbol\theta}}_
{i}}\boldsymbol\Gamma_i\boldsymbol\tau_{i,n}$\\
$\quad+\sum\limits_{k = 1}^{n - 2}\frac{\partial{\alpha}_{i,k}}{\partial{\hat{\boldsymbol\theta}}_
{i}}\boldsymbol\Gamma_i\boldsymbol\omega_{i,n}e_{i,k+1}$\\
with\\$\boldsymbol\omega_{i,n}=\boldsymbol\varphi_{i,n}-\sum\nolimits_{j = 1}^{n - 1} \frac{\partial{\alpha}_{i,n-1}}{\partial{x}_{i,j}}\boldsymbol\varphi_{i,j}$\\$\boldsymbol\tau_{i,n}=\boldsymbol
\tau_{i,n-1}+\boldsymbol\omega_{i,n}e_{i,n}$
} & (T1.3)
\\ \hline
\tabincell{l}{Parameter Update Laws:\\$\dot{\chi}_i=e_{i,n}\bar{u}_i$\\
$\dot{\hat{\bm{\theta}}}_i=\boldsymbol\Gamma_i\boldsymbol\tau_{i,n}$} & (T1.4)
\\ \hline
\end{tabular}
\end{table}

\textbf{\emph{Step $k ~(k=2,...,n)$}}: For brevity, the control design details of remaining steps are summarized in Table I with $c_{i,k}$ being positive constant. Compared with \emph{Step} 2, the extra term $\sum\nolimits_{j = 2}^{k - 1}\frac{\partial{\alpha}_{i,j-1}}{\partial{\hat{\boldsymbol\theta}}_
{i}}\boldsymbol\Gamma_i\boldsymbol\omega_{i,k}e_{i,j}$ in (T1.2) is used for counteracting the residual term in previous step.

\subsubsection{Stability Analysis}
The main result of the adaptive backstepping tracking control algorithm can be summarized in the following theorem.
\begin{theorem}\label{Theorem 2}
Consider the closed-loop system consisting of $N$ agents \eqref{eq.15}, the filters \eqref{eq.1}, the control laws (T1.3) with adaptive laws (T1.4) under \emph{Assumptions} \ref{Assumption 1}--\ref{Assumption 2}. For any given initial condition satisfying that $\zeta_{\varepsilon_i}(0) \in \mathcal{B}_{r_0}$, then all the closed-loop signals are SGUUB and $y_i(t)$ asymptotically tracks $\hat y_i(t)$ with arbitrarily prescribed transient performance.
\end{theorem}

\textbf{\emph{Proof}}:
Choose the quadratic Lyapunov function as
\begin{align}\label{eq.58}
V_{i,n}=V_{i,1}+\sum\limits_{k = 2}^{n}\frac{1}{2}e^2_{i,k}.
\end{align}
From \eqref{eq.21}, \eqref{eq.25} and (T.1)--(T.3), the derivative of $V_{i,n}$ is
\begin{align}\label{eq.36}
\dot{V}_{i,n}&=\left( g_i(\boldsymbol x_i, t)N_i(\chi_i)+1 \right)e_{i,n}{\bar u}_i
+\tilde{\boldsymbol\theta}^T_i\boldsymbol\Gamma^{-1}_i\left(\boldsymbol\Gamma_i\boldsymbol\tau_{i,n}-\dot{\hat{\boldsymbol\theta}}_i\right) \nonumber\\
&-\sum\limits_{k = 1}^{n}c_{i,k}e^2_{i,k}
-\sum\limits_{k=1}^{n-1}\frac{\partial{\alpha}_{i,k}}{\partial{\hat{\boldsymbol\theta}}_{i}}\left(\dot{\hat{\boldsymbol\theta}}_i
-\boldsymbol\Gamma_i\boldsymbol\tau_{i,n}\right)e_{i,k+1}
\end{align}
In view of (T1.4), we arrive at
\begin{align}\label{eq.37}
\dot{V}_{i,n}=&\left(g_i(\boldsymbol x_i, t)N_i(\chi_i)+1 \right)\dot{\chi}_i-\sum\limits_{k = 1}^{n}c_{i,k}e^2_{i,k} \nonumber\\
\leq & \left(g_i(\boldsymbol x_i, t)N_i(\chi_i)+1 \right)\dot{\chi}_i.
\end{align}
According to \emph{Lemma} \ref{Lemma 4}, it is established that $V_{i,n}(t)$ and $\chi_i(t)$ are bounded on $[0,\infty)$, then the boundedness of $e_{i,k} ~(k=1,...,n)$ and $\hat{\boldsymbol\theta}_i$ is ensured from the definition of $V_{i,n}$ along with \eqref{eq.40} and \eqref{eq.58}. With the boundedness of $e_{i,1}$, it is deduced from \eqref{eq.48} that $\xi_i$ is bounded, which then implies that $|\varepsilon_i(t)|<\beta_{i2}(t)$, $\forall t \geq0$. From the definition of $\mu_i$, it follows that $\mu_i$ is bounded. Moreover, since $\hat{y}_i$ is bounded, the boundedness of $x_{i,1}$ is also guaranteed. Subsequently, $\alpha_{i,1}$ and $\dot{e}_{i,1}$ are bounded from the smoothness of $\boldsymbol\varphi_{i,k}$. Similarly, the boundedness of $x_{i,k}$, $\alpha_{i,j}$, $\dot{e}_{i,k}$ ~$(k=2,...,n; j=2,...,n-1)$ and $u_i$ can also be established. Hence, all signals in the closed-loop system are bounded.


Next, by integrating both sides of \eqref{eq.37}, we have
\begin{align}
V_{i,n}(t)=&-\sum\limits_{k=1}^n c_{i,k} \int_0^te^2_{i,k}(\tau)d\tau + V_{i,n}(0) \nonumber\\
&+ \int_0^t \left( g_i(\boldsymbol x_i, t)N_i(\chi_i)+1 \right)\dot{\chi}_i(\tau)d\tau,
\end{align}
which means that $e_{i,k} \in L_2$. Based on the Barbalat's Lemma \cite{khalil2002nonlinear}, it further derives that $\lim _{{t} \to +\infty } e_{i,k}(t) = 0$, and thus it is obtained from \eqref{eq.16}--\eqref{eq.48} that $\lim _{{t} \to +\infty } \varepsilon_i(t) = 0$. Hence, the output of each agent tracks the corresponding filtered signal asymptotically, where the transient performance can be arbitrarily prescribed by selecting the parameters $\iota$ and $\beta_{i,2}$ properly. \hfill $\blacksquare$ 
\subsection{Final Result}
The prescribed performance of the filter error $\delta_i(t)$ and the auxiliary tracking error $\varepsilon_i(t)$ is ensured in Sections III-A and III-B, respectively. Hence, the final result of this paper can be summarized in the following theorem.
\begin{theorem}\label{Theorem 3}
For uncertain nonlinear MASs \eqref{eq.15} under \emph{Assumptions} \ref{Assumption 1}--\ref{Assumption 2}, the control schemes \eqref{eq.7}, (T1.3) and (T1.4) solve the prescribed performance control
problem in the presence of unknown time-varying coefficients, i.e., the output tracking error of each agent converges into an arbitrarily pre-assigned residual set exhibiting an arbitrarily pre-defined convergence rate.
\end{theorem}

\textbf{\emph{Proof}}: With the distributed control law \eqref{eq.7}, it is derived from \emph{Theorem} \ref{Theorem 1} that \eqref{eq.43} (equivalent to \eqref{eq.49}) is satisfied, then the performance bounds of $\delta_i$ can be deduced by \eqref{eq.42}. In view of (T1.3) and (T1.4), the pre-specified performance \eqref{eq.19} of $\varepsilon_i$ is ensured from \emph{Theorem} \ref{Theorem 2}. Hence, the prescribed performance of output tracking error $e_i$ can be guaranteed indirectly with arbitrarily pre-defined converge speed and arbitrarily pre-assigned size of the residual set.
\hfill $\blacksquare$

\begin{remark}\label{Remark 7}
It has been shown that the prescribed output tracing performance for each agent is achieved via an indirect framework that is comprised of distributed robust filter and adaptive backstepping control scheme. Compared with the existing method, some notable advantages of ours can be summarized as follows:
\begin{itemize}
\item \emph{Distributed prescribed performance tracking}: Under the proposed indirect framework, the distributed performance-guaranteed tracking control is achieved in the sense that the output tracking error converges into an arbitrarily pre-defined residual set with an arbitrarily pre-assignable convergence rate, which thus is more powerful than the methods in \cite{psillakis2016consensus, wang2018cooperative, wang2020adaptive, huang2018fully, wang2020adaptiveb}, since, on one hand, only the leaderless consensus is achieved in \cite{psillakis2016consensus, wang2018cooperative}; on the other hand, the transient performance/steady state behavior in \cite{psillakis2016consensus, wang2018cooperative, wang2020adaptive, huang2018fully, wang2020adaptiveb} cannot be quantificationally guaranteed.
\item \emph{Lower complexity}: In contrast to \cite{cai2015leader, huang2017smooth, chen2015observer, liu2017adaptive}, the co-design control scheme remains at a lower structural complexity since the distributed robust filter is free from any adaptive parameters to be estimated online. In addition, only a set of uniform Nussbaum functions are used to deal with the unknown and non-identical control direction problem, which significantly simplifies the control design and the stability analysis as compared with \cite{wang2020adaptive, huang2018fully, wang2020adaptiveb} since different and complicated Nussbaum functions are required therein.

\item \emph{Broader scope of application}: By developing the novel \emph{Lemma} \ref{Lemma 4} regarding Nussbaum function, the unknown control coefficients in \eqref{eq.15} are relaxed to be time-varying and even state-dependent, which greatly broadens the scope of application of the proposed algorithms since the control coefficients are limited to be constant in \cite{chen2013adaptive, wang2020adaptive, liu2014adaptive, ding2015adaptive, psillakis2016consensus, wang2018cooperative} and the time-varying and state-dependent term is required to be known in \cite{huang2018fully, wang2020adaptiveb}.
\end{itemize}
\end{remark}


\begin{remark}\label{Remark 9}
Note that the boundedness of $O_i$ and $\xi_i$ is sufficient to ensure the desired output tracking performance as long as certain initial conditions, i.e., $\zeta_{\bar{z}_i}(0) \in \mathcal{B}_{r_0}$ and $\zeta_{\varepsilon_i}(0) \in \mathcal{B}_{r_0}$ are satisfied. In this sense, the design recipe of the parameters is as follows:
\begin{itemize}
\item  The convergence speed of the filter error $\delta_i$ and the auxiliary tracking error $\varepsilon_i$ is dictated by the parameter $\iota$, i.e., larger $\iota$ leads to faster convergence speed of $\delta_i$ and $\varepsilon_i$.

\item  The steady-state error of $\delta_i$ and $\varepsilon_i$ is dominated by the parameters $\beta_{1,\infty} ~(\rho_\infty)$ and $\beta_{2,\infty}$, i.e., smaller $\beta_{1,\infty}$ and $\beta_{2,\infty}$ result in smaller allowable error of $\delta_i$ and $\varepsilon_i$ at steady-state, respectively.

\item Larger $c_{i,k}$ helps to improve the tracking performance (but not influencing the performance metrics), i.e., making the trajectory smoother within prescribed performance bounds, which however could result in higher control efforts. Therefore, $c_{i,k}$ is not necessary to be chosen too large.
\end{itemize}
\end{remark}




\section{Simulation Studies}
To verify the effectiveness of the proposed scheme, we now consider a group of four nonlinear agents with the following dynamics
\begin{align}\label{eq.41}
\dot x_{i,1}&=x_{i,2}+\theta^T_i\varphi_{i,1}(x_{i,1}), \nonumber\\
\dot x_{i,2}&=g_i(x_i,t)u_i+\theta^T_i\varphi_{i,2}(x_i),
\end{align}
where the smooth nonlinear functions are set as $\varphi_{i,1}(\cdot)={\rm sin}(x_{i,1}) ~(i=1,...,4)$, $\varphi_{1,2}(\cdot)=x_{1,2}$, $\varphi_{2,2}(\cdot)=x_{2,2}$, $\varphi_{3,2}(\cdot)=x_{3,1}x_{3,2}$, $\varphi_{4,2}(\cdot)=x_{4,1}x_{4,2}$; the unknown parameters are chosen as $\theta_1=0.7$, $\theta_2=0.8$, $\theta_3=0.5$, $\theta_4=0.6$. The communication topology is given in Fig. \ref{Fig.2} such that \emph{Assumption} \ref{Assumption 1} is satisfied. Since the signs of the control coefficients are allowed to be unknown and non-identical, two cases are respectively considered as follows:

\begin{figure}[t]
  \centering
\includegraphics[height=0.19\textwidth]{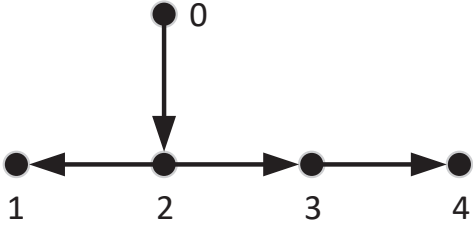}
  \caption{The communication topology.}
  \label{Fig.2}
\end{figure}

\begin{figure}[htp]
 \centering
\includegraphics[height=0.26\textwidth]{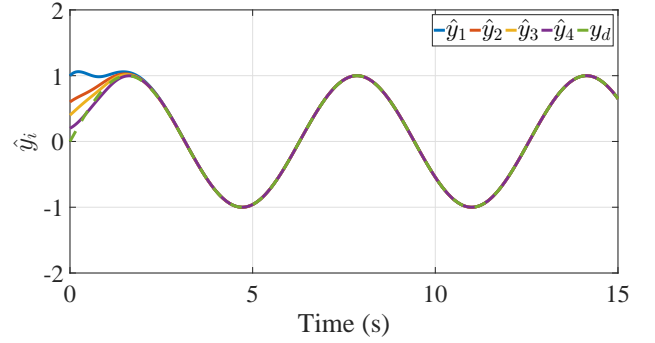}
  \subfigure{(a) Trajectory of $\hat{y}_i$.}
\includegraphics[height=0.26\textwidth]{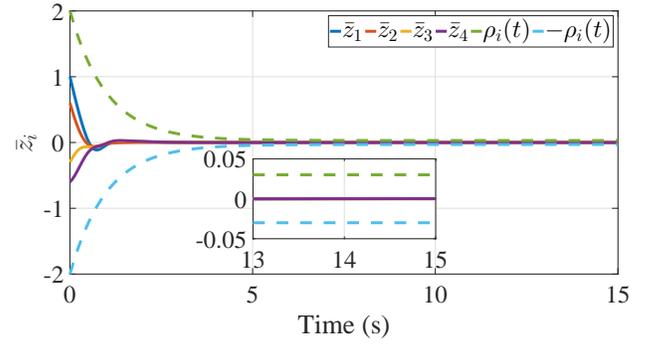}
   \subfigure{(b) Trajectory of $\bar z_i$.}
\includegraphics[height=0.26\textwidth]{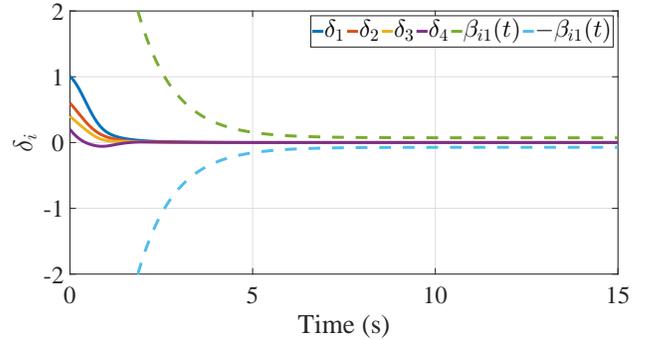}
  \subfigure{(c) Trajectory of $\delta_i$.}
\includegraphics[height=0.26\textwidth]{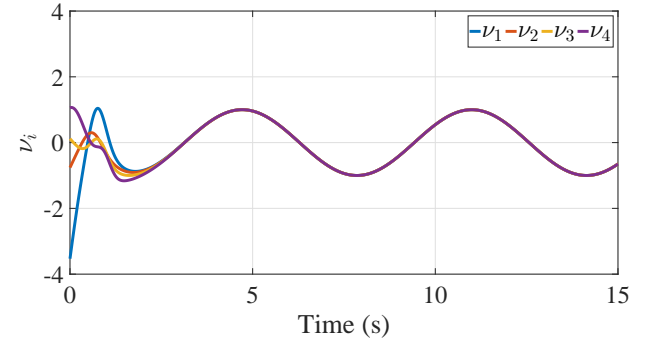}
  \subfigure{(d) Trajectory of $\nu_i$.}
   \caption{Response of the filter under \eqref{eq.7}.}
   \label{Fig.3}
 \end{figure}

  \begin{figure}[htp]
 \centering
  \includegraphics[height=0.26\textwidth]{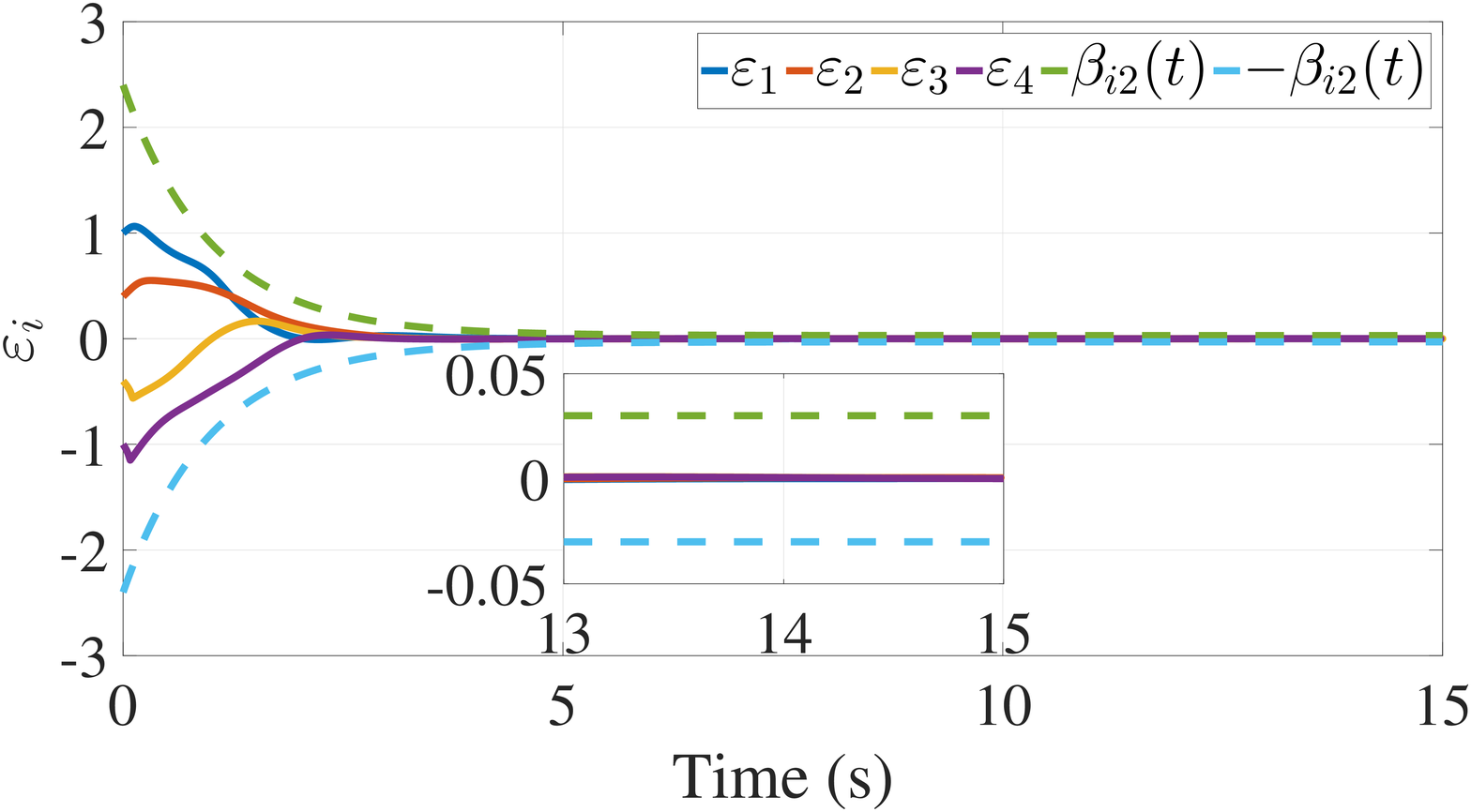}
  \subfigure{(a) Trajectory of $\varepsilon_i$.}
  \includegraphics[height=0.26\textwidth]{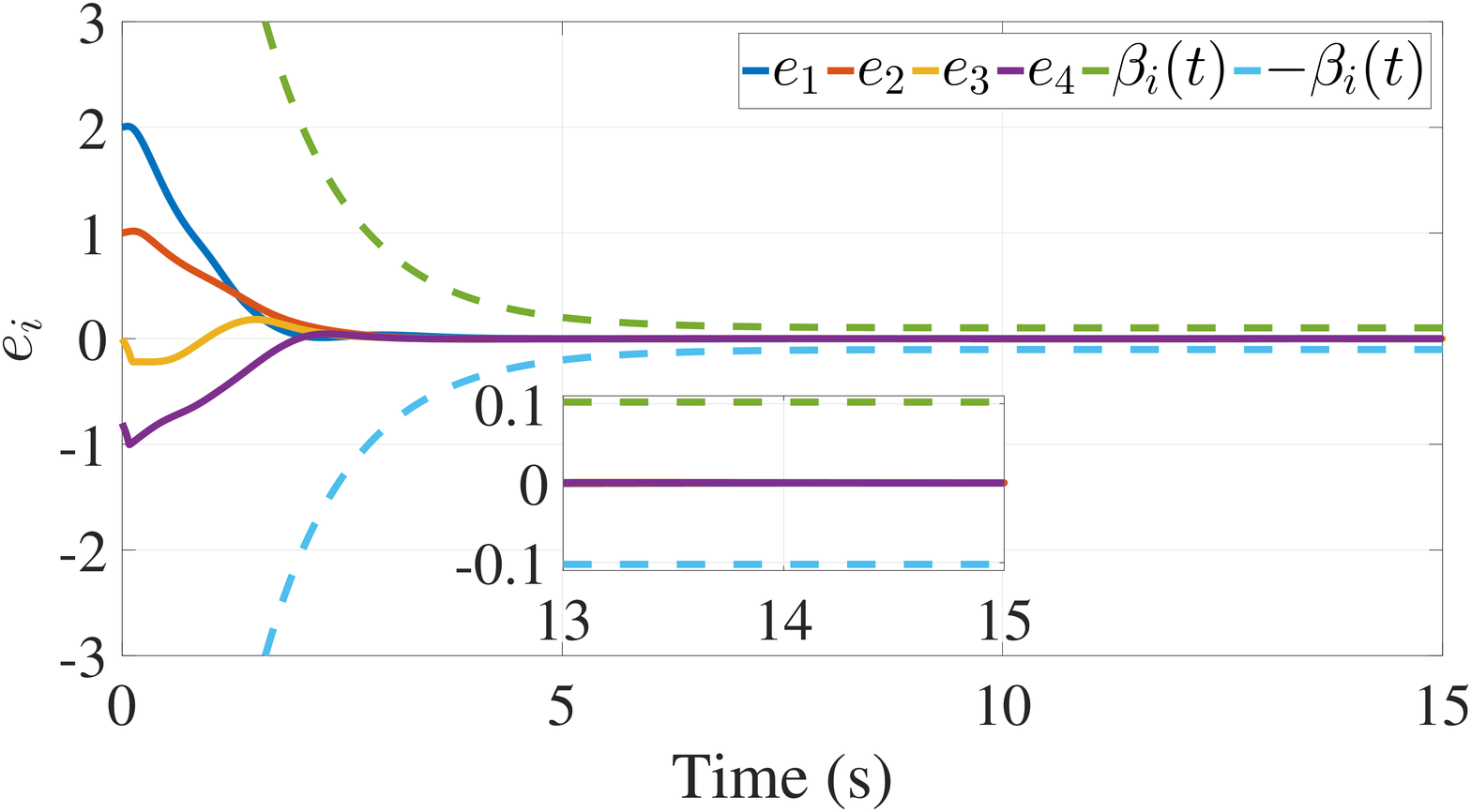}
   \subfigure{(b) Trajectory of $e_i$.}
   \includegraphics[height=0.26\textwidth]{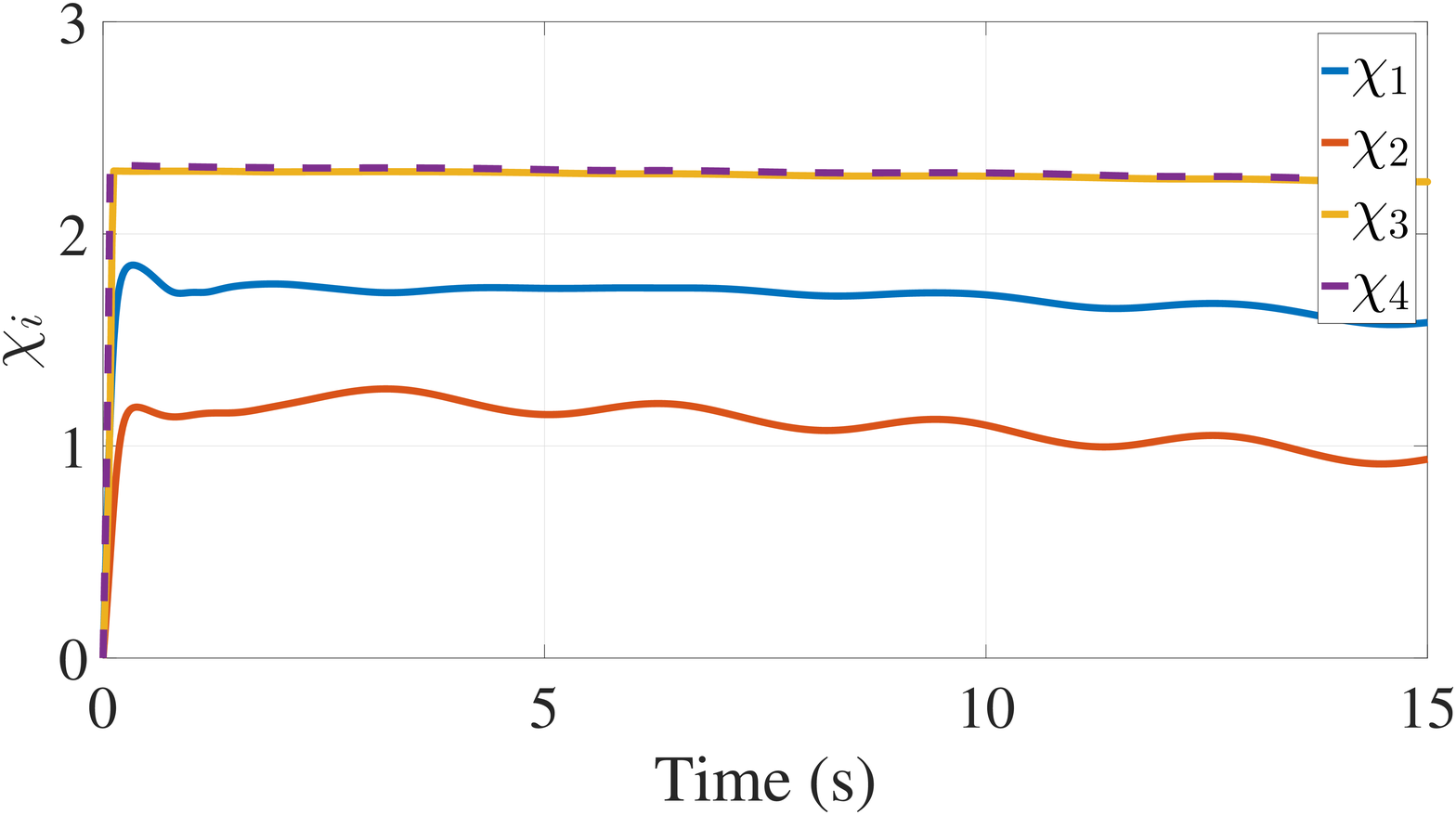}
   \subfigure{(c) Trajectory of $\chi_i$.}
   \includegraphics[height=0.26\textwidth]{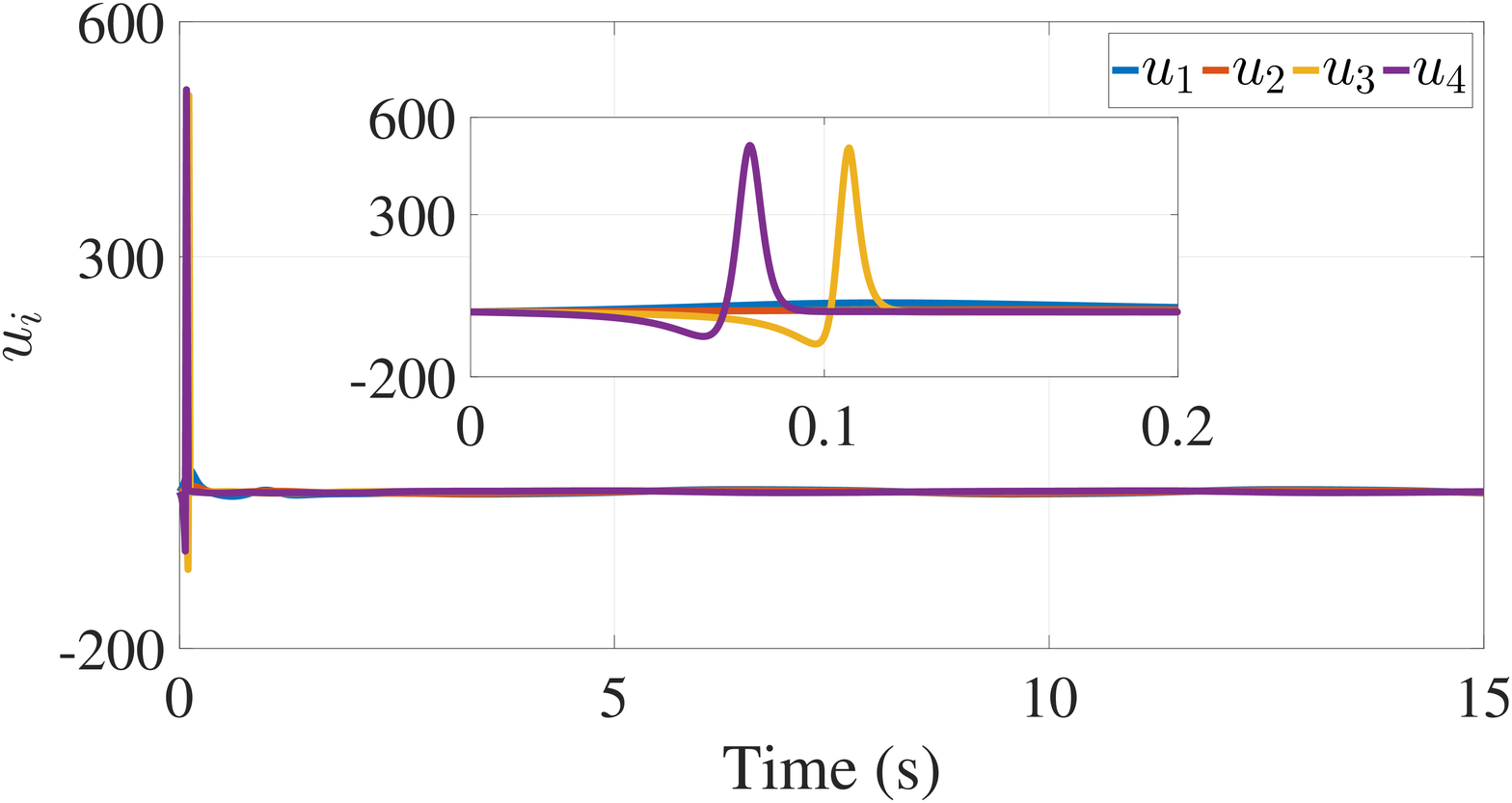}
  \subfigure{(d) Trajectory of $u_i$.}
  \caption{Response of controller under (T1.3) and (T1.4) with $g_i(\cdot)$.}
   \label{Fig.4}
\end{figure}

\begin{figure}[htp]
 \centering
  \includegraphics[height=0.26\textwidth]{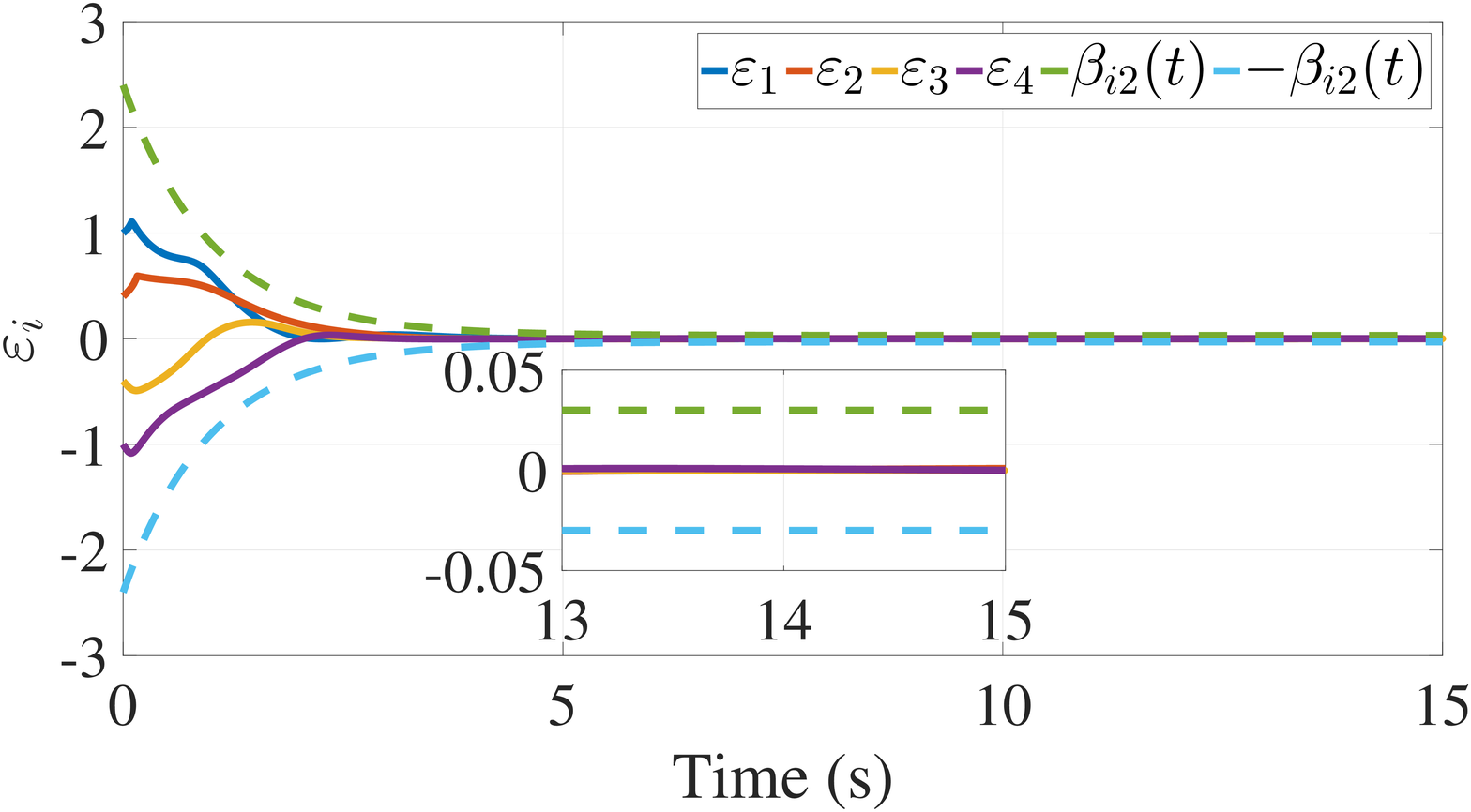}
  \subfigure{(a) Trajectory of $\varepsilon_i$.}
  \includegraphics[height=0.26\textwidth]{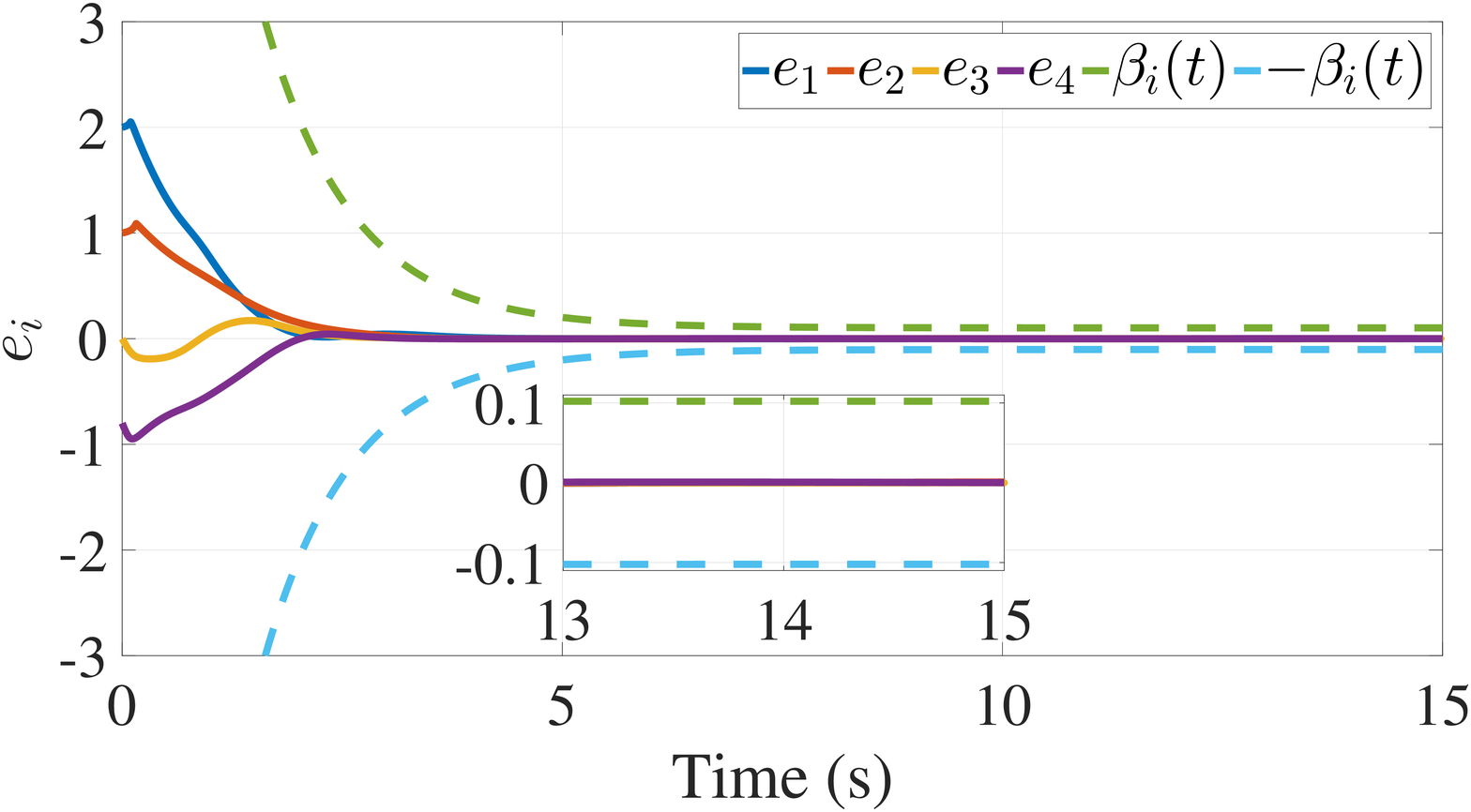}
   \subfigure{(b) Trajectory of $e_i$.}
     \includegraphics[height=0.26\textwidth]{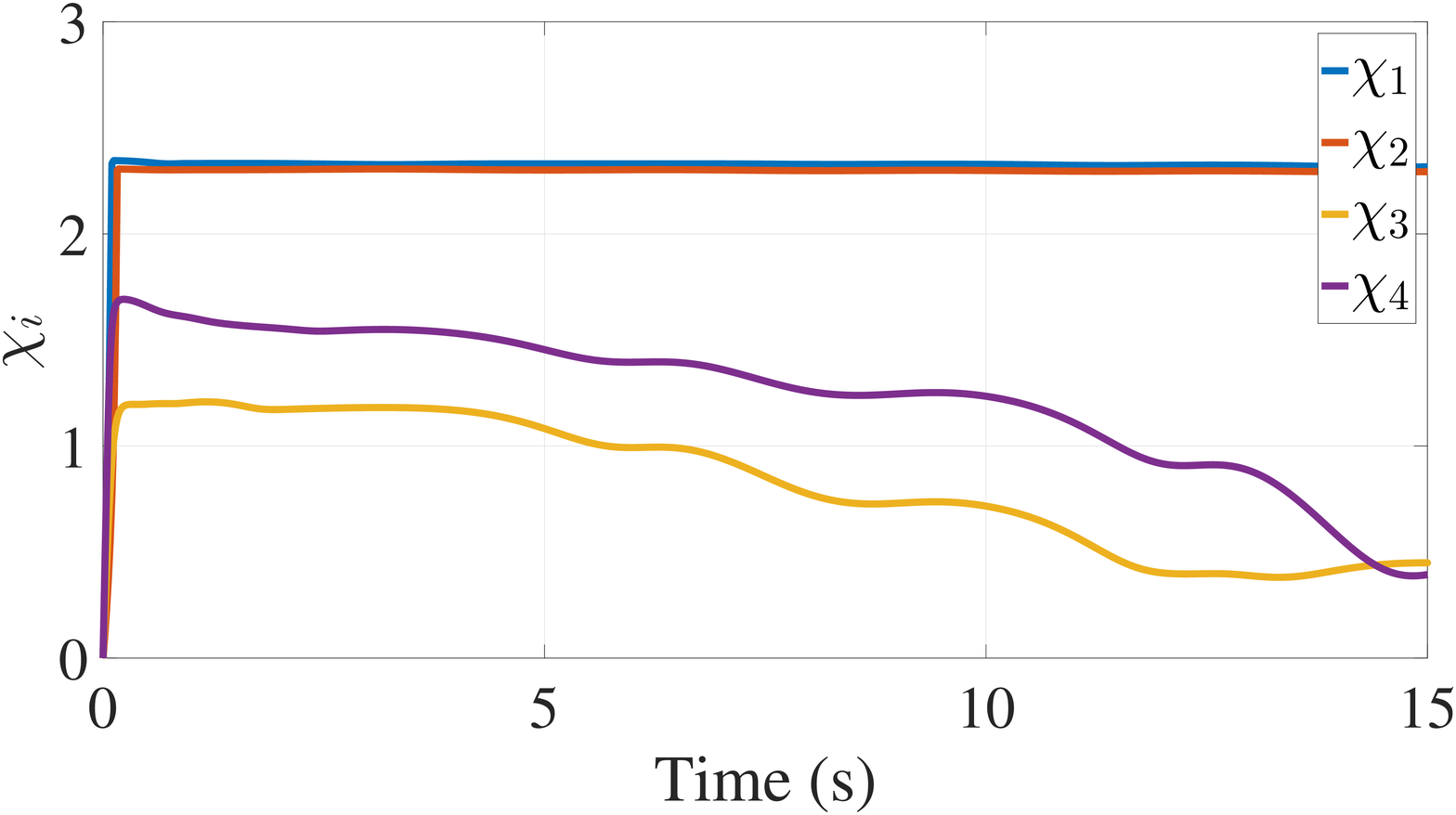}
   \subfigure{(c) Trajectory of $\chi_i$.}
   \includegraphics[height=0.26\textwidth]{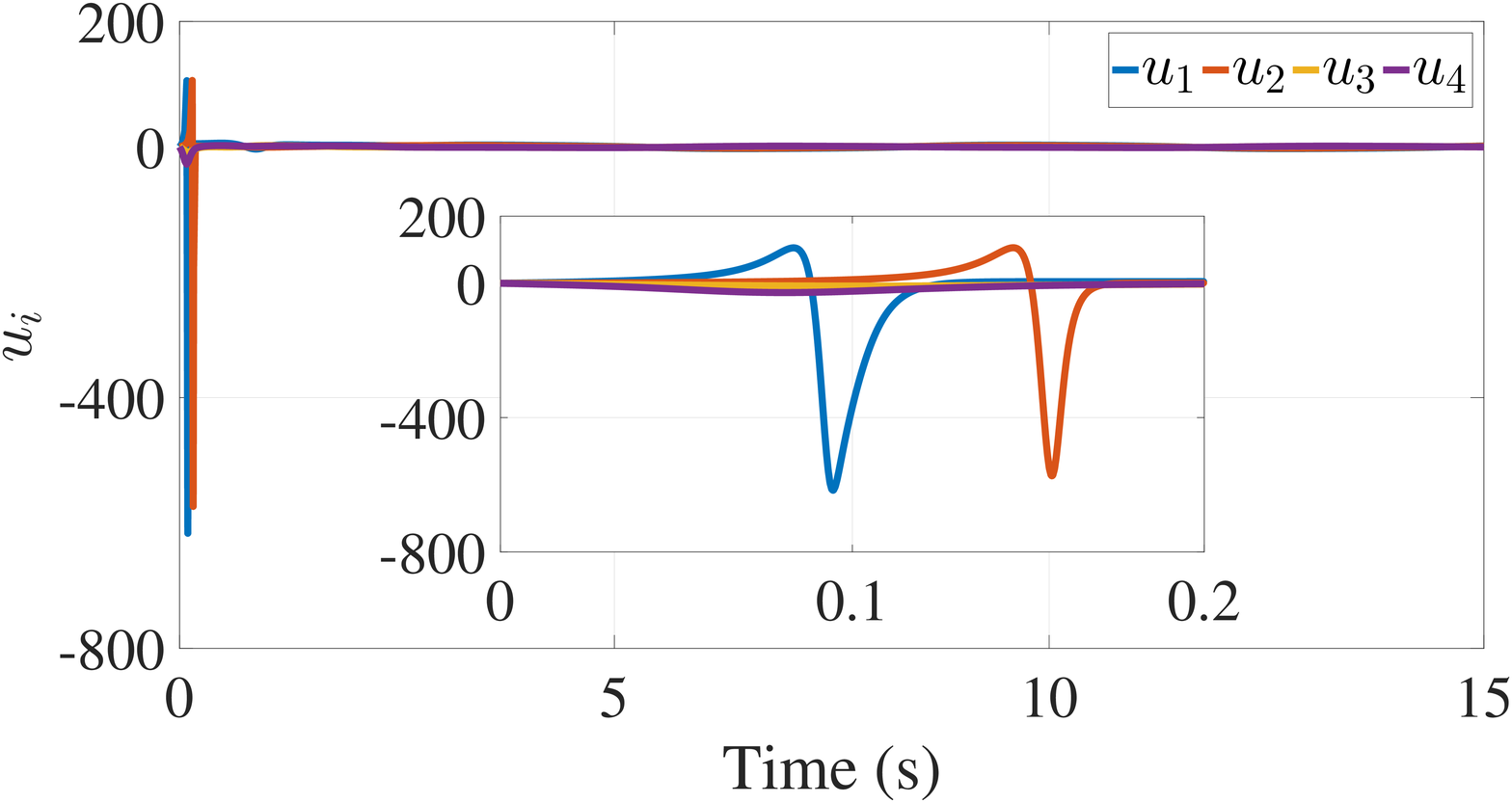}
  \subfigure{(d) Trajectory of $u_i$.}
   \caption{Response of controller under (T1.3) and (T1.4) with $g'_i(\cdot)$.}
   \label{Fig.5}
\end{figure}

\begin{itemize}
\item[\textbf{1) :}]
$g_1(\cdot)=-0.5-0.1{\rm sin}(x_{1,1}x_{1,2})$, $g_2(\cdot)=-0.6-0.2{\rm cos}(x_{2,1}x_{2,2})$, $g_3(\cdot)=1+0.2{\rm cos}(x^2_{3,2})$, $g_4(\cdot)=1+0.1{\rm sin}(x^2_{4,1})$;
\item[\textbf{2) :}]
$g'_i(\cdot)=-g_i(\cdot) ~(i=1,...,4)$.
\end{itemize}

\noindent In both cases, the signs of $g_i$ are non-identical, which satisfies the conditions imposed in \emph{Assumption} \ref{Assumption 2}.

In the simulation, the desired trajectory is set as $y_0(t)={\rm sin}(t)$. The performance functions for $\bar z_i$ and $\varepsilon_i$ are prespecified as $\rho_{i}(t)=1.97{\rm exp}(-\iota t)+0.03$ and $\beta_{i2}(t)=2.37{\rm exp}(-\iota t)+0.03 ~(i=1,...,4)$ with $\iota=1$, respectively, which means that: 1) the steady-state error of $\bar z_i$ and $\varepsilon_i$ is smaller than 0.03; and 2) the convergence rates of them are faster than ${\rm exp}(-\iota t)$. By \eqref{eq.22}, \eqref{eq.42} and \eqref{eq.19}, the performance function for $\delta_i$ and $e_i$ are calculated as $\beta_{i1}(t)=12.3617{\rm exp}(-\iota t)+0.0722$, $\beta_i(t)=14.7317{\rm exp}(-\iota t)+0.1022$. The initial states of filters and agents are set as $\hat {y}_{i}(0)=[1, 0.6, 0.4, 0.2]^T$, $\dot{\hat y}_{i}(0)=[0.6, 0.4, 0.5, 0.3]^T$, $\bar {x}_{i,1}(0)=[2, 1, 0, -0.8]^T$, $\bar {x}_{i,2}(0)=[0.6, 0.5, -0.5, -0.6]^T$, the rest is set to be zero. The design parameters are selected as $\lambda=2$ in \eqref{eq.3} and \eqref{eq.4}, $c_i=4$ in \eqref{eq.7}, $\bar {c}_{i,1}=[1, 2, 1.5, 1.2]^T$ in \eqref{eq.23}, $\bar {c}_{i,2}=[1, 2, 1.5, 1.2]^T$ in (T1.3), $\bar \Gamma_i=[0.5, 0.1, 2, 0.5]^T$ in (T1.4). The Nussbaum functions in (T1.3) are uniformly chosen as $N_i(\chi_i)=e^{\chi^2_i}{\rm sin}(\chi_i\pi/2) ~(i=1,...,4)$.

The simulation results of \emph{Case} 1 are shown in Fig. \ref{Fig.3} and Fig. \ref{Fig.4}. Fig. \ref{Fig.3} is the response of the designed performance-guaranteed filter under \eqref{eq.7}. It can be seen from Fig. \ref{Fig.3} (a)--(c) that the prescribed performance of $\bar z_i$ and $\delta_i$ is achieved, and the filter input $\nu_i$ in Fig. \ref{Fig.3} (d) is bounded. Fig. \ref{Fig.4} is the response of the system under (T1.3) and (T1.4), which indicates that both the auxiliary tracking error $\varepsilon_i$ in Fig. \ref{Fig.4} (a) and the output tracking error $e_i$ in Fig. \ref{Fig.4} (b) are preserving within the predefined bounds under unknown non-identical control directions and time-varying coefficients. Additionally, the actual input signal $u_i$ in Fig. \ref{Fig.4} (d) is bounded.

To further verify the efficacy of developed Nussbaum-based scheme, the scenario in \emph{Case} 2 is also test with the initial conditions and parameters selection remaining unchanged. The corresponding simulation results
are depicted in Fig. \ref{Fig.5}. It is shown that the prescribed tracking performance is also satisfactorily ensured, which is consistent with the theoretical findings.
Moreover, the striking difference between Fig. \ref{Fig.4} and Fig. \ref{Fig.5} lies in the amplitude of the input signal $u_i$ for each agent, which implies that the unknown control directions can be accurately identified with our method whether it is positive or negative.

Different from \cite{chen2013adaptive, wang2020adaptive, liu2014adaptive, ding2015adaptive, psillakis2016consensus, wang2018cooperative, huang2018fully, wang2020adaptiveb}, in which only one unknown control direction is considered for each agent, we conduct the simulation by using two opposite control directions in this paper to verify the validity of the proposed method adequately.

\section{Conclusion}
This paper studies the distributed prescribed performance control of networked uncertain strict-feedback MASs with time-varying gains in a leader-following scenario. A new yet indirect performance-guaranteed framework that combines distributed robust filters with backstepping adaptive control design is developed, of which not only the unknown control directions are allowed to be non-identical, but also the control coefficients are permitted be to time-varying and even state-dependent by establishing a novel lemma regarding Nussbaum function. What is more, the performance of output tracking error is guaranteed with arbitrarily pre-assignable converge rate and arbitrarily prescribed size of the residual set, which is independent of the underlying graph topology and can be explicitly determined by the design parameters. Considering the uncertain MASs with more general forms, e.g., pure feedback systems \cite{huang2020output}, within this framework represents a future research point.

\section*{Appendix}\label{Appendix}
\begin{figure}[tbp]
  \centering
  \includegraphics[height=0.19\textwidth]{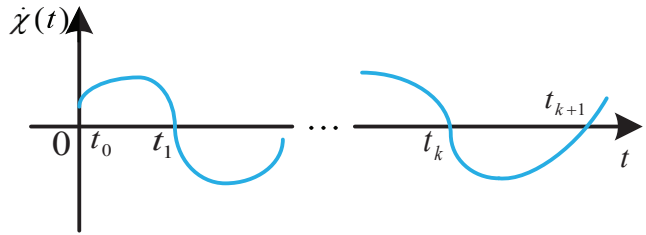}
  \caption{The division of time domain: $\dot{\chi} \geq 0$, for $t \in [t_0, t_1)$ and  $\dot{\chi} \leq 0$, for $t \in [t_k, t_{k+1})$.}
  \label{Fig.7}
\end{figure}

\setcounter{equation}{0}
\renewcommand\theequation{A.\arabic{equation}}
\textbf{\emph{A.}} \textbf{\emph{Proof of Lemma \ref{Lemma 4}}}:
The proof begins with the division of the entire time domain into infinite continuous time intervals, i.e., $[t_k,t_{k+1})$, $k=0, ... , \infty$, as shown in Fig. \ref{Fig.7}. For convenience and clarity, the boundedness of $\chi(t)$ and $V(t)$ over $[t_0, \infty)$ is proved by mathematical induction, which consists of three steps as follows.

\textbf{\emph{Step I}}: We show that $\chi(t)$ and $V(t)$ are bounded for all $t \in [t_0,t_1)$. Let
\begin{align}
\hat N(s)=\eta_2N^+(s)-\eta_1N^-(s), \quad\eta_1\eta_2>0. \nonumber
\end{align}
According to \cite[\emph{Lemma} 4.2]{chen2019nussbaum}, $\hat N(s)$ is also a Nussbaum function (type B-$\hat K$) with
\begin{align}
\hat K = {\rm min} \left\{ \frac{\eta_2}{\eta_1}, \frac{\eta_1}{\eta_2}\right\}K > {\rm min} \left\{ \frac{\eta_2}{\eta_1}, \frac{\eta_1}{\eta_2}\right\} {\rm max} \left\{ \frac{\eta_2}{\eta_1}, \frac{\eta_1}{\eta_2}\right\} =1. \nonumber
\end{align}
From \eqref{eq.44}, it follows that $N(s)=N^+(s)-N^-(s)$. Then, it is further derived for all $\tau \in [t_0, t_1)$ that
\begin{align}\label{eq.46}
\eta(\tau)N(\chi(\tau))&=\eta(\tau)N^+(\chi(\tau))-\eta(\tau)N^-(\chi(\tau)) \nonumber\\
& \leq \eta_2N^+(\chi(\tau))-\eta_1N^-(\chi(\tau)) \nonumber\\
&=\hat {N}(\chi(\tau)).
\end{align}
Integrating both sides of \eqref{eq.17} over $t \in [t_0, t_1)$ along with \eqref{eq.46} yields,
\begin{align}\label{eq.18}
&V(t) \leq \int_{t_0}^{t} \left[{\eta(\tau)} N(\chi(\tau))+a\right]\dot{\chi}(\tau)d\tau + V(t_0) \nonumber\\
&= \int_{t_0}^{t} \eta(\tau)N(\chi(\tau))d(\chi(\tau)) + a\int_{t_0}^{t} \dot{\chi}(\tau)d\tau + V(t_0) \nonumber\\
&\leq \int_{\chi(t_0)}^{\chi(t)} \hat{N}(s)ds + a\int_{t_0}^{t} \dot{\chi}(\tau)d\tau + V(t_0) \nonumber\\
&= \int_0^{\chi(t)} \hat{N}(s)ds - \int_0^{\chi(t_0)} \hat{N}(s)ds + a\chi(t) - a\chi(t_0) + V(t_0) \nonumber\\
&= \int_0^{\chi(t)} \hat{N}(s)ds + a\chi(t) - c_0,
\end{align}
where $c_0=\int_0^{\chi(t_0)} \hat{N}(s)ds + a\chi(t_0) - V(t_0)$ is a constant. Considering the fact that $V(t)\geq 0$, it is further deduced from \eqref{eq.18} that
\begin{align}\label{eq.47}
\int_0^{\chi(t)} \hat{N}(s)ds + a\chi(t) \geq c_0, \quad \forall t \in [t_0, t_1).
\end{align}
At this stage, two cases need to be discussed, respectively.

\begin{figure}
  \centering
  \includegraphics[height=0.2\textwidth]{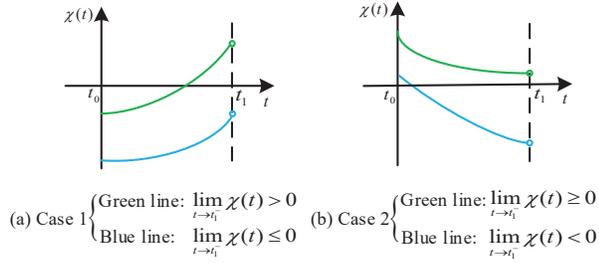}
\caption[flushleft]{The possible evolution of $\chi(t)$ over $[t_0, t_1)$.}
  \label{Fig.8}
\end{figure}

\textbf{\emph{Case 1}}: When $\dot{\chi}(t) \geq 0$ for all $t \in [t_0, t_1)$. Since $\hat{N}(\chi)$ is a traditional Nussbaum function (type A) defined as in \cite{xudong1998adaptive}, there exists a constant $\chi_1^*>1$ such that
\begin{align}
\frac{1}{\chi_1^*} \int_0^{\chi_1^*} \hat{N}(s)ds < -\left| c_0 \right|-a. \nonumber
\end{align}
Next, two scenarios, as illustrated in Fig. \ref{Fig.8} (a), should be considered as follows.

\emph{\underline{Scenario 1--a}: ${\lim _{t \to t_1^ - }}\chi (t) > 0$ (green line).}
In this situation, the boundedness of $\chi(t)$ and $V(t)$ over $[t_0, t_1)$ is proved by contradiction. Firstly, it is supposed that $\chi(t)$ is unbounded over $[t_0, t_1)$, there must exist a time instant $t_1^* \in [t_0, t_1)$ such that $\chi(t_1^*)=\chi_1^*$, then
\begin{align}
\frac{1}{\chi(t_1^*)} \int_0^{\chi(t_1^*)} \hat{N}(s)ds < -\left| c_0 \right|-a < \frac{c_0}{\chi(t_1^*)}-a. \nonumber
\end{align}
Subsequently,
\begin{align}
\int_0^{\chi(t_1^*)} \hat{N}(s)ds + a\chi(t_1^*)< c_0, \nonumber
\end{align}
which leads to a contradiction with \eqref{eq.47}. Hence $\chi(t)$ is bounded over $[t_0, t_1)$, so is $V(t)$ based on \eqref{eq.18}. 

\emph{\underline{Scenario 1--b}: ${\lim _{t \to t_1^ - }}\chi (t) \leq 0$ (blue line).}
The boundedness of $\chi(t)$ over $[t_0, t_1)$ is immediate due to the boundedness of $\chi(t_0)$, ${\lim _{t \to t_1^ - }}\chi (t) \leq 0$ and $\dot{\chi}(t) \geq 0$ over $[t_0, t_1)$. Then it follows from \eqref{eq.18} that $V(t)$ is also bounded $\forall t \in [t_0, t_1)$.

\textbf{\emph{Case 2}}: When $\dot{\chi}(t) < 0$ for all $t \in [t_0, t_1)$. Similar to \emph{Case} 1, for the Nussbaum function $\hat{N}(\chi)$, there exists a constant $\chi_2^*<-1$ such that
\begin{align}
\frac{1}{\chi_2^*} \int_0^{\chi_2^*} \hat{N}(s)ds > \left| c_0 \right|-a. \nonumber
\end{align}
Then two scenarios, as illustrated in Fig. \ref{Fig.8} (b), should be discussed as follows.

\emph{\underline{Scenario 2--a}: ${\lim _{t \to t_1^ - }}\chi (t) < 0$ (blue line).}
Analogous to \emph{Scenario 1-a}, proof by contradiction is utilized. Suppose that $\chi(t)$ is unbounded over $[t_0, t_1)$, a time instant $t_2^* \in [t_0, t_1)$ must exist such that $\chi(t_2^*)=\chi_2^*$, then
\begin{align}
\frac{1}{\chi(t_2^*)} \int_0^{\chi(t_2^*)} \hat{N}(s)ds > \left| c_0 \right|-a >\frac{c_0}{\chi(t_2^*)}-a. \nonumber
\end{align}
Consequently, it is deduced that
\begin{align}
\int_0^{\chi(t_2^*)} \hat{N}(s)ds + a\chi(t_2^*)< c_0, \nonumber
\end{align}
which contradicts \eqref{eq.47}. Thus, $\chi(t)$ is bounded over $[t_0, t_1)$, and the same conclusion for $V(t)$ is obtained from \eqref{eq.18}.

\emph{\underline{Scenario 2--b}: ${\lim _{t \to t_1^ - }}\chi (t) \geq 0$ (green line).}
Since $\chi(t_0)$ is bounded and $\dot{\chi}(t) < 0$ over $[t_0, t_1)$ with $\chi(t_1) \geq 0$, the boundedness of $\chi(t)$ and $V(t)$ over $[t_0, t_1)$ can be ensured.

\textbf{\emph{Step II}}: Suppose $\chi(t)$ and $V(t)$ are bounded over $[t_{k-1}, t_k)$, $k=2,..., \infty$.

\textbf{\emph{Step III}}: In view of the continuity of $\chi(t)$ and $V(t)$, it holds that
\begin{align}
\chi ({t_k}) = \mathop {\lim }\limits_{t \to t_k^ - } \chi (t), \quad V ({t_k}) = \mathop {\lim }\limits_{t \to t_k^ - } V (t), \nonumber
\end{align}
which implies that $\chi(t_k)$ and $V(t_k)$ are bounded under the assumption in \emph{Step II}. By following the same reasoning procedure as in \emph{Step I}, it is derived that the boundedness of $\chi(t)$ and $V(t)$ over $[t_k, t_{k+1})$ can be guaranteed. This completes the proof.
\hfill $\blacksquare$

\textbf{\emph{B.}} \textbf{\emph{Proof of Proposition 1}}:
Consider the following $k$ first-order linear low pass filters with $\omega_k(t)$ being the output and pole $-\lambda$ driven by the scalar quantity $\bar z_i(t)$, i.e.,
\begin{align}
\bar z_i(t)=\left(\frac{d}{dt}+\lambda\right)^k\omega_k(t). \nonumber
\end{align}
It can be readily obtained for $k=1$ that
\begin{align}
|\omega_1(t)| \leq |\omega_1(0)|e^{-\lambda t}+\int_0^t e^{-\lambda(t-\tau)}|\bar z_i(\tau)|d\tau. \nonumber
\end{align}
Since $|\bar z_i(t)| < \rho_{i}(t)$, $\forall t \geq 0$, and $\lambda > \iota$, it follows that
\begin{align}
|\omega_1(t)| \leq &|\omega_1(0)|e^{-\iota t}+ \rho_{\infty} \int_0^t e^{-\lambda(t-\tau)}d\tau \nonumber\\
&+(\rho_{0}-\rho_{\infty})e^{-\lambda t} \int_0^t e^{(\lambda-\iota)\tau}d\tau \nonumber\\
\leq & |\omega_1(0)|e^{-\iota t}+\frac{\rho_{\infty}}{\lambda}
+\frac{\rho_{0}-\rho_{\infty}}{\lambda-\iota} e^{-\iota t}  \nonumber\\
=& \bar \omega_1 e^{-\iota t}+\frac{\rho_{\infty}}{\lambda}, \nonumber
\end{align}
with $\bar \omega_1=|\omega_1(0)|+(\rho_{0}-\rho_{\infty})/(\lambda-\iota)$ being a positive constant. By carrying out the same recursively reasoning, it is concluded that
\begin{align}
|\omega_k(t)| \leq \bar \omega_k e^{-\iota t}+\frac{\rho_{\infty}}{\lambda^k}, \quad k=1,...,n-1 \nonumber
\end{align}
with $\bar \omega_k= |\omega_k(0)|+\bar \omega_{k-1}/(\lambda-\iota), ~k=2,...,n-1$. Note from \eqref{eq.4} that $\omega_{n-1}(t)=z_i(t)$, we have
\begin{align}
|z_i(t)| \leq \bar \omega_{n-1} e^{-\iota t}+\frac{\rho_{\infty}}{\lambda^{n-1}}, \quad i=1,...,N \nonumber
\end{align}
then
\begin{align}
\|\boldsymbol z\| \leq \sqrt{N}\left(\bar \omega_{n-1} e^{-\iota t}+\frac{\rho_{\infty}}{\lambda^{n-1}}\right), \nonumber
\end{align}
which, along with \eqref{eq.55}, implies that
\begin{align}\label{eq.42}
\|\boldsymbol\delta\|
\leq \frac{\sqrt{N}\left(\bar \omega_{n-1} e^{-\iota t}+\frac{\rho_{\infty}}{\lambda^{n-1}}\right)}{\sigma_{\rm min}(\boldsymbol{ L}+\boldsymbol{ B})}=\beta_{i1}(t)
\end{align}
for all $t \geq 0$. Thus,
\begin{align}
\lim _{{t} \to +\infty } \|\boldsymbol\delta\| = \frac{\sqrt{N}\rho_{\infty}}{\lambda^{n-1}\sigma_{\rm min}(\boldsymbol{ L}+\boldsymbol{ B})}, \nonumber
\end{align}
which completes the proof.
\hfill $\blacksquare$

\bibliographystyle{IEEEtran}
\bibliography{filter_PPC_2022}

 \end{document}